\documentclass[longauth]{aa}

\usepackage{graphicx}
\usepackage{natbib}
\usepackage{scalerel}
\usepackage{xcolor}
\usepackage[normalem]{ulem}  % Matthias: allows crossing out text
\usepackage{orcidlink}
\usepackage{euclid}

\definecolor{purple1}{rgb}{0.5,0,0.87}

\usepackage{txfonts}
\hypersetup{
    colorlinks=true,
    linkcolor=blue,
    filecolor=magenta,      
    urlcolor=blue,
    citecolor=blue,
    pdftitle={Euclid: Early Release Observations – The intracluster light of Abell 2390},
    pdfauthor={}
}
\makeatletter
\renewcommand*\aa@pageof{, page \thepage{} of \pageref*{LastPage}}
\makeatother

\usepackage[utf8]{inputenc}

\usepackage[switch, modulo]{lineno}
\usepackage{euclid}

\begin{document}
%
% Put the title and authors of your (Standard Project) paper here
%

\title{\Euclid: Early Release Observations -  The intracluster light of Abell~2390 \thanks{This paper is published on behalf of the Euclid Consortium.}}    
%% please do not edit the author list once you copy it from the
%% Publication Portal -- contact ECEB Bureau for changes
%%%% please do not edit the author list -- contact ECEB Bureau for changes
\newcommand{\orcid}[1]{\orcidlink{#1}} %% if already defined in aa.cls: comment, or use renewcommand			   
%%%% please do not edit the author list -- contact ECEB Bureau for changes
\author{A.~Ellien\orcid{0000-0002-1038-3370}\thanks{\email{amael.ellien@oca.eu}}\inst{\ref{aff1}}
\and M.~Montes\orcid{0000-0001-7847-0393}\inst{\ref{aff2},\ref{aff3},\ref{aff4}}
\and S.~L.~Ahad\orcid{0000-0001-6336-642X}\inst{\ref{aff5},\ref{aff6}}
\and P.~Dimauro\orcid{0000-0001-7399-2854}\inst{\ref{aff7},\ref{aff8}}
\and J.~B.~Golden-Marx\orcid{0000-0002-6394-045X}\inst{\ref{aff9}}
\and Y.~Jimenez-Teja\orcid{0000-0002-6090-2853}\inst{\ref{aff10},\ref{aff8}}
\and F.~Durret\orcid{0000-0002-6991-4578}\inst{\ref{aff11}}
\and C.~Bellhouse\orcid{0000-0002-6179-8007}\inst{\ref{aff9}}
\and J.~M.~Diego\orcid{0000-0001-9065-3926}\inst{\ref{aff12}}
\and S.~P.~Bamford\orcid{0000-0001-7821-7195}\inst{\ref{aff9}}
\and A.~H.~Gonzalez\orcid{0000-0002-0933-8601}\inst{\ref{aff13}}
\and N.~A.~Hatch\orcid{0000-0001-5600-0534}\inst{\ref{aff9}}
\and M.~Kluge\orcid{0000-0002-9618-2552}\inst{\ref{aff14}}
\and R.~Ragusa\inst{\ref{aff15}}
\and E.~Slezak\orcid{0000-0003-4771-7263}\inst{\ref{aff16}}
\and J.-C.~Cuillandre\orcid{0000-0002-3263-8645}\inst{\ref{aff17}}
\and R.~Gavazzi\orcid{0000-0002-5540-6935}\inst{\ref{aff18},\ref{aff19}}
\and H.~Dole\orcid{0000-0002-9767-3839}\inst{\ref{aff20}}
\and G.~Mahler\orcid{0000-0003-3266-2001}\inst{\ref{aff21},\ref{aff22},\ref{aff23}}
\and G.~Congedo\orcid{0000-0003-2508-0046}\inst{\ref{aff24}}
\and T.~Saifollahi\orcid{0000-0002-9554-7660}\inst{\ref{aff25}}
\and N.~Aghanim\orcid{0000-0002-6688-8992}\inst{\ref{aff20}}
\and B.~Altieri\orcid{0000-0003-3936-0284}\inst{\ref{aff26}}
\and A.~Amara\inst{\ref{aff27}}
\and S.~Andreon\orcid{0000-0002-2041-8784}\inst{\ref{aff28}}
\and N.~Auricchio\orcid{0000-0003-4444-8651}\inst{\ref{aff29}}
\and C.~Baccigalupi\orcid{0000-0002-8211-1630}\inst{\ref{aff30},\ref{aff31},\ref{aff32},\ref{aff33}}
\and M.~Baldi\orcid{0000-0003-4145-1943}\inst{\ref{aff34},\ref{aff29},\ref{aff35}}
\and A.~Balestra\orcid{0000-0002-6967-261X}\inst{\ref{aff36}}
\and S.~Bardelli\orcid{0000-0002-8900-0298}\inst{\ref{aff29}}
\and A.~Basset\inst{\ref{aff37}}
\and P.~Battaglia\orcid{0000-0002-7337-5909}\inst{\ref{aff29}}
\and A.~Biviano\orcid{0000-0002-0857-0732}\inst{\ref{aff31},\ref{aff30}}
\and A.~Bonchi\orcid{0000-0002-2667-5482}\inst{\ref{aff38}}
\and D.~Bonino\orcid{0000-0002-3336-9977}\inst{\ref{aff39}}
\and E.~Branchini\orcid{0000-0002-0808-6908}\inst{\ref{aff40},\ref{aff41},\ref{aff28}}
\and M.~Brescia\orcid{0000-0001-9506-5680}\inst{\ref{aff42},\ref{aff15},\ref{aff43}}
\and J.~Brinchmann\orcid{0000-0003-4359-8797}\inst{\ref{aff44},\ref{aff45}}
\and A.~Caillat\inst{\ref{aff18}}
\and S.~Camera\orcid{0000-0003-3399-3574}\inst{\ref{aff46},\ref{aff47},\ref{aff39}}
\and V.~Capobianco\orcid{0000-0002-3309-7692}\inst{\ref{aff39}}
\and C.~Carbone\orcid{0000-0003-0125-3563}\inst{\ref{aff48}}
\and V.~F.~Cardone\inst{\ref{aff7},\ref{aff49}}
\and J.~Carretero\orcid{0000-0002-3130-0204}\inst{\ref{aff50},\ref{aff51}}
\and S.~Casas\orcid{0000-0002-4751-5138}\inst{\ref{aff52},\ref{aff53}}
\and M.~Castellano\orcid{0000-0001-9875-8263}\inst{\ref{aff7}}
\and G.~Castignani\orcid{0000-0001-6831-0687}\inst{\ref{aff29}}
\and S.~Cavuoti\orcid{0000-0002-3787-4196}\inst{\ref{aff15},\ref{aff43}}
\and A.~Cimatti\inst{\ref{aff54}}
\and C.~Colodro-Conde\inst{\ref{aff3}}
\and C.~J.~Conselice\orcid{0000-0003-1949-7638}\inst{\ref{aff55}}
\and L.~Conversi\orcid{0000-0002-6710-8476}\inst{\ref{aff56},\ref{aff26}}
\and Y.~Copin\orcid{0000-0002-5317-7518}\inst{\ref{aff57}}
\and F.~Courbin\orcid{0000-0003-0758-6510}\inst{\ref{aff58},\ref{aff59}}
\and H.~M.~Courtois\orcid{0000-0003-0509-1776}\inst{\ref{aff60}}
\and M.~Cropper\orcid{0000-0003-4571-9468}\inst{\ref{aff61}}
\and A.~Da~Silva\orcid{0000-0002-6385-1609}\inst{\ref{aff62},\ref{aff63}}
\and H.~Degaudenzi\orcid{0000-0002-5887-6799}\inst{\ref{aff64}}
\and G.~De~Lucia\orcid{0000-0002-6220-9104}\inst{\ref{aff31}}
\and A.~M.~Di~Giorgio\orcid{0000-0002-4767-2360}\inst{\ref{aff65}}
\and J.~Dinis\orcid{0000-0001-5075-1601}\inst{\ref{aff62},\ref{aff63}}
\and F.~Dubath\orcid{0000-0002-6533-2810}\inst{\ref{aff64}}
\and C.~A.~J.~Duncan\orcid{0009-0003-3573-0791}\inst{\ref{aff55}}
\and X.~Dupac\inst{\ref{aff26}}
\and S.~Dusini\orcid{0000-0002-1128-0664}\inst{\ref{aff66}}
\and M.~Farina\orcid{0000-0002-3089-7846}\inst{\ref{aff65}}
\and F.~Faustini\orcid{0000-0001-6274-5145}\inst{\ref{aff38},\ref{aff7}}
\and S.~Ferriol\inst{\ref{aff57}}
\and S.~Fotopoulou\orcid{0000-0002-9686-254X}\inst{\ref{aff67}}
\and M.~Frailis\orcid{0000-0002-7400-2135}\inst{\ref{aff31}}
\and E.~Franceschi\orcid{0000-0002-0585-6591}\inst{\ref{aff29}}
\and S.~Galeotta\orcid{0000-0002-3748-5115}\inst{\ref{aff31}}
\and K.~George\orcid{0000-0002-1734-8455}\inst{\ref{aff68}}
\and B.~Gillis\orcid{0000-0002-4478-1270}\inst{\ref{aff24}}
\and C.~Giocoli\orcid{0000-0002-9590-7961}\inst{\ref{aff29},\ref{aff35}}
\and P.~G\'omez-Alvarez\orcid{0000-0002-8594-5358}\inst{\ref{aff69},\ref{aff26}}
\and A.~Grazian\orcid{0000-0002-5688-0663}\inst{\ref{aff36}}
\and F.~Grupp\inst{\ref{aff14},\ref{aff68}}
\and L.~Guzzo\orcid{0000-0001-8264-5192}\inst{\ref{aff70},\ref{aff28}}
\and S.~V.~H.~Haugan\orcid{0000-0001-9648-7260}\inst{\ref{aff71}}
\and J.~Hoar\inst{\ref{aff26}}
\and H.~Hoekstra\orcid{0000-0002-0641-3231}\inst{\ref{aff72}}
\and W.~Holmes\inst{\ref{aff73}}
\and F.~Hormuth\inst{\ref{aff74}}
\and A.~Hornstrup\orcid{0000-0002-3363-0936}\inst{\ref{aff75},\ref{aff76}}
\and P.~Hudelot\inst{\ref{aff19}}
\and K.~Jahnke\orcid{0000-0003-3804-2137}\inst{\ref{aff77}}
\and M.~Jhabvala\inst{\ref{aff78}}
\and B.~Joachimi\orcid{0000-0001-7494-1303}\inst{\ref{aff79}}
\and E.~Keih\"anen\orcid{0000-0003-1804-7715}\inst{\ref{aff80}}
\and S.~Kermiche\orcid{0000-0002-0302-5735}\inst{\ref{aff81}}
\and A.~Kiessling\orcid{0000-0002-2590-1273}\inst{\ref{aff73}}
\and B.~Kubik\orcid{0009-0006-5823-4880}\inst{\ref{aff57}}
\and K.~Kuijken\orcid{0000-0002-3827-0175}\inst{\ref{aff72}}
\and M.~K\"ummel\orcid{0000-0003-2791-2117}\inst{\ref{aff68}}
\and M.~Kunz\orcid{0000-0002-3052-7394}\inst{\ref{aff82}}
\and H.~Kurki-Suonio\orcid{0000-0002-4618-3063}\inst{\ref{aff83},\ref{aff84}}
\and R.~Laureijs\inst{\ref{aff85},\ref{aff86}}
\and D.~Le~Mignant\orcid{0000-0002-5339-5515}\inst{\ref{aff18}}
\and S.~Ligori\orcid{0000-0003-4172-4606}\inst{\ref{aff39}}
\and P.~B.~Lilje\orcid{0000-0003-4324-7794}\inst{\ref{aff71}}
\and V.~Lindholm\orcid{0000-0003-2317-5471}\inst{\ref{aff83},\ref{aff84}}
\and I.~Lloro\orcid{0000-0001-5966-1434}\inst{\ref{aff87}}
\and G.~Mainetti\orcid{0000-0003-2384-2377}\inst{\ref{aff88}}
\and D.~Maino\inst{\ref{aff70},\ref{aff48},\ref{aff89}}
\and E.~Maiorano\orcid{0000-0003-2593-4355}\inst{\ref{aff29}}
\and O.~Mansutti\orcid{0000-0001-5758-4658}\inst{\ref{aff31}}
\and S.~Marcin\inst{\ref{aff90}}
\and O.~Marggraf\orcid{0000-0001-7242-3852}\inst{\ref{aff91}}
\and K.~Markovic\orcid{0000-0001-6764-073X}\inst{\ref{aff73}}
\and M.~Martinelli\orcid{0000-0002-6943-7732}\inst{\ref{aff7},\ref{aff49}}
\and N.~Martinet\orcid{0000-0003-2786-7790}\inst{\ref{aff18}}
\and F.~Marulli\orcid{0000-0002-8850-0303}\inst{\ref{aff92},\ref{aff29},\ref{aff35}}
\and R.~Massey\orcid{0000-0002-6085-3780}\inst{\ref{aff23}}
\and S.~Maurogordato\inst{\ref{aff16}}
\and E.~Medinaceli\orcid{0000-0002-4040-7783}\inst{\ref{aff29}}
\and S.~Mei\orcid{0000-0002-2849-559X}\inst{\ref{aff93},\ref{aff94}}
\and M.~Melchior\inst{\ref{aff95}}
\and Y.~Mellier\inst{\ref{aff11},\ref{aff19}}
\and M.~Meneghetti\orcid{0000-0003-1225-7084}\inst{\ref{aff29},\ref{aff35}}
\and E.~Merlin\orcid{0000-0001-6870-8900}\inst{\ref{aff7}}
\and G.~Meylan\inst{\ref{aff96}}
\and A.~Mora\orcid{0000-0002-1922-8529}\inst{\ref{aff97}}
\and M.~Moresco\orcid{0000-0002-7616-7136}\inst{\ref{aff92},\ref{aff29}}
\and L.~Moscardini\orcid{0000-0002-3473-6716}\inst{\ref{aff92},\ref{aff29},\ref{aff35}}
\and R.~Nakajima\orcid{0009-0009-1213-7040}\inst{\ref{aff91}}
\and C.~Neissner\orcid{0000-0001-8524-4968}\inst{\ref{aff98},\ref{aff51}}
\and R.~C.~Nichol\orcid{0000-0003-0939-6518}\inst{\ref{aff27}}
\and S.-M.~Niemi\inst{\ref{aff85}}
\and J.~W.~Nightingale\orcid{0000-0002-8987-7401}\inst{\ref{aff99}}
\and C.~Padilla\orcid{0000-0001-7951-0166}\inst{\ref{aff98}}
\and S.~Paltani\orcid{0000-0002-8108-9179}\inst{\ref{aff64}}
\and F.~Pasian\orcid{0000-0002-4869-3227}\inst{\ref{aff31}}
\and K.~Pedersen\inst{\ref{aff100}}
\and W.~J.~Percival\orcid{0000-0002-0644-5727}\inst{\ref{aff5},\ref{aff6},\ref{aff101}}
\and V.~Pettorino\inst{\ref{aff85}}
\and S.~Pires\orcid{0000-0002-0249-2104}\inst{\ref{aff17}}
\and G.~Polenta\orcid{0000-0003-4067-9196}\inst{\ref{aff38}}
\and M.~Poncet\inst{\ref{aff37}}
\and L.~A.~Popa\inst{\ref{aff102}}
\and L.~Pozzetti\orcid{0000-0001-7085-0412}\inst{\ref{aff29}}
\and F.~Raison\orcid{0000-0002-7819-6918}\inst{\ref{aff14}}
\and R.~Rebolo\inst{\ref{aff3},\ref{aff103},\ref{aff4}}
\and A.~Renzi\orcid{0000-0001-9856-1970}\inst{\ref{aff104},\ref{aff66}}
\and J.~Rhodes\orcid{0000-0002-4485-8549}\inst{\ref{aff73}}
\and G.~Riccio\inst{\ref{aff15}}
\and E.~Romelli\orcid{0000-0003-3069-9222}\inst{\ref{aff31}}
\and M.~Roncarelli\orcid{0000-0001-9587-7822}\inst{\ref{aff29}}
\and E.~Rossetti\orcid{0000-0003-0238-4047}\inst{\ref{aff34}}
\and R.~Saglia\orcid{0000-0003-0378-7032}\inst{\ref{aff68},\ref{aff14}}
\and Z.~Sakr\orcid{0000-0002-4823-3757}\inst{\ref{aff105},\ref{aff106},\ref{aff107}}
\and D.~Sapone\orcid{0000-0001-7089-4503}\inst{\ref{aff108}}
\and B.~Sartoris\orcid{0000-0003-1337-5269}\inst{\ref{aff68},\ref{aff31}}
\and R.~Scaramella\orcid{0000-0003-2229-193X}\inst{\ref{aff7},\ref{aff49}}
\and M.~Schirmer\orcid{0000-0003-2568-9994}\inst{\ref{aff77}}
\and P.~Schneider\orcid{0000-0001-8561-2679}\inst{\ref{aff91}}
\and T.~Schrabback\orcid{0000-0002-6987-7834}\inst{\ref{aff109}}
\and A.~Secroun\orcid{0000-0003-0505-3710}\inst{\ref{aff81}}
\and E.~Sefusatti\orcid{0000-0003-0473-1567}\inst{\ref{aff31},\ref{aff30},\ref{aff32}}
\and G.~Seidel\orcid{0000-0003-2907-353X}\inst{\ref{aff77}}
\and M.~Seiffert\orcid{0000-0002-7536-9393}\inst{\ref{aff73}}
\and S.~Serrano\orcid{0000-0002-0211-2861}\inst{\ref{aff110},\ref{aff111},\ref{aff2}}
\and C.~Sirignano\orcid{0000-0002-0995-7146}\inst{\ref{aff104},\ref{aff66}}
\and G.~Sirri\orcid{0000-0003-2626-2853}\inst{\ref{aff35}}
\and L.~Stanco\orcid{0000-0002-9706-5104}\inst{\ref{aff66}}
\and J.-L.~Starck\orcid{0000-0003-2177-7794}\inst{\ref{aff17}}
\and J.~Steinwagner\orcid{0000-0001-7443-1047}\inst{\ref{aff14}}
\and P.~Tallada-Cresp\'{i}\orcid{0000-0002-1336-8328}\inst{\ref{aff50},\ref{aff51}}
\and A.~N.~Taylor\inst{\ref{aff24}}
\and H.~I.~Teplitz\orcid{0000-0002-7064-5424}\inst{\ref{aff112}}
\and I.~Tereno\inst{\ref{aff62},\ref{aff113}}
\and R.~Toledo-Moreo\orcid{0000-0002-2997-4859}\inst{\ref{aff114}}
\and F.~Torradeflot\orcid{0000-0003-1160-1517}\inst{\ref{aff51},\ref{aff50}}
\and A.~Tsyganov\inst{\ref{aff115}}
\and I.~Tutusaus\orcid{0000-0002-3199-0399}\inst{\ref{aff106}}
\and L.~Valenziano\orcid{0000-0002-1170-0104}\inst{\ref{aff29},\ref{aff116}}
\and T.~Vassallo\orcid{0000-0001-6512-6358}\inst{\ref{aff68},\ref{aff31}}
\and G.~Verdoes~Kleijn\orcid{0000-0001-5803-2580}\inst{\ref{aff86}}
\and A.~Veropalumbo\orcid{0000-0003-2387-1194}\inst{\ref{aff28},\ref{aff41},\ref{aff40}}
\and Y.~Wang\orcid{0000-0002-4749-2984}\inst{\ref{aff112}}
\and J.~Weller\orcid{0000-0002-8282-2010}\inst{\ref{aff68},\ref{aff14}}
\and O.~R.~Williams\orcid{0000-0003-0274-1526}\inst{\ref{aff115}}
\and E.~Zucca\orcid{0000-0002-5845-8132}\inst{\ref{aff29}}
\and M.~Bolzonella\orcid{0000-0003-3278-4607}\inst{\ref{aff29}}
\and C.~Burigana\orcid{0000-0002-3005-5796}\inst{\ref{aff117},\ref{aff116}}
\and V.~Scottez\inst{\ref{aff11},\ref{aff118}}}
										   
%%%% please do not edit the affiliation list -- contact ECEB Bureau for changes
\institute{OCA, P.H.C Boulevard de l'Observatoire CS 34229, 06304 Nice Cedex 4, France\label{aff1}
\and
Institute of Space Sciences (ICE, CSIC), Campus UAB, Carrer de Can Magrans, s/n, 08193 Barcelona, Spain\label{aff2}
\and
Instituto de Astrof\'{\i}sica de Canarias, V\'{\i}a L\'actea, 38205 La Laguna, Tenerife, Spain\label{aff3}
\and
Universidad de La Laguna, Departamento de Astrof\'{\i}sica, 38206 La Laguna, Tenerife, Spain\label{aff4}
\and
Waterloo Centre for Astrophysics, University of Waterloo, Waterloo, Ontario N2L 3G1, Canada\label{aff5}
\and
Department of Physics and Astronomy, University of Waterloo, Waterloo, Ontario N2L 3G1, Canada\label{aff6}
\and
INAF-Osservatorio Astronomico di Roma, Via Frascati 33, 00078 Monteporzio Catone, Italy\label{aff7}
\and
Observatorio Nacional, Rua General Jose Cristino, 77-Bairro Imperial de Sao Cristovao, Rio de Janeiro, 20921-400, Brazil\label{aff8}
\and
School of Physics and Astronomy, University of Nottingham, University Park, Nottingham NG7 2RD, UK\label{aff9}
\and
Instituto de Astrof\'isica de Andaluc\'ia, CSIC, Glorieta de la Astronom\'\i a, 18080, Granada, Spain\label{aff10}
\and
Institut d'Astrophysique de Paris, 98bis Boulevard Arago, 75014, Paris, France\label{aff11}
\and
Instituto de F\'isica de Cantabria, Edificio Juan Jord\'a, Avenida de los Castros, 39005 Santander, Spain\label{aff12}
\and
Department of Astronomy, University of Florida, Bryant Space Science Center, Gainesville, FL 32611, USA\label{aff13}
\and
Max Planck Institute for Extraterrestrial Physics, Giessenbachstr. 1, 85748 Garching, Germany\label{aff14}
\and
INAF-Osservatorio Astronomico di Capodimonte, Via Moiariello 16, 80131 Napoli, Italy\label{aff15}
\and
Universit\'e C\^{o}te d'Azur, Observatoire de la C\^{o}te d'Azur, CNRS, Laboratoire Lagrange, Bd de l'Observatoire, CS 34229, 06304 Nice cedex 4, France\label{aff16}
\and
Universit\'e Paris-Saclay, Universit\'e Paris Cit\'e, CEA, CNRS, AIM, 91191, Gif-sur-Yvette, France\label{aff17}
\and
Aix-Marseille Universit\'e, CNRS, CNES, LAM, Marseille, France\label{aff18}
\and
Institut d'Astrophysique de Paris, UMR 7095, CNRS, and Sorbonne Universit\'e, 98 bis boulevard Arago, 75014 Paris, France\label{aff19}
\and
Universit\'e Paris-Saclay, CNRS, Institut d'astrophysique spatiale, 91405, Orsay, France\label{aff20}
\and
STAR Institute, Quartier Agora - All\'ee du six Ao\^ut, 19c B-4000 Li\`ege, Belgium\label{aff21}
\and
Department of Physics, Centre for Extragalactic Astronomy, Durham University, South Road, Durham, DH1 3LE, UK\label{aff22}
\and
Department of Physics, Institute for Computational Cosmology, Durham University, South Road, Durham, DH1 3LE, UK\label{aff23}
\and
Institute for Astronomy, University of Edinburgh, Royal Observatory, Blackford Hill, Edinburgh EH9 3HJ, UK\label{aff24}
\and
Universit\'e de Strasbourg, CNRS, Observatoire astronomique de Strasbourg, UMR 7550, 67000 Strasbourg, France\label{aff25}
\and
ESAC/ESA, Camino Bajo del Castillo, s/n., Urb. Villafranca del Castillo, 28692 Villanueva de la Ca\~nada, Madrid, Spain\label{aff26}
\and
School of Mathematics and Physics, University of Surrey, Guildford, Surrey, GU2 7XH, UK\label{aff27}
\and
INAF-Osservatorio Astronomico di Brera, Via Brera 28, 20122 Milano, Italy\label{aff28}
\and
INAF-Osservatorio di Astrofisica e Scienza dello Spazio di Bologna, Via Piero Gobetti 93/3, 40129 Bologna, Italy\label{aff29}
\and
IFPU, Institute for Fundamental Physics of the Universe, via Beirut 2, 34151 Trieste, Italy\label{aff30}
\and
INAF-Osservatorio Astronomico di Trieste, Via G. B. Tiepolo 11, 34143 Trieste, Italy\label{aff31}
\and
INFN, Sezione di Trieste, Via Valerio 2, 34127 Trieste TS, Italy\label{aff32}
\and
SISSA, International School for Advanced Studies, Via Bonomea 265, 34136 Trieste TS, Italy\label{aff33}
\and
Dipartimento di Fisica e Astronomia, Universit\`a di Bologna, Via Gobetti 93/2, 40129 Bologna, Italy\label{aff34}
\and
INFN-Sezione di Bologna, Viale Berti Pichat 6/2, 40127 Bologna, Italy\label{aff35}
\and
INAF-Osservatorio Astronomico di Padova, Via dell'Osservatorio 5, 35122 Padova, Italy\label{aff36}
\and
Centre National d'Etudes Spatiales -- Centre spatial de Toulouse, 18 avenue Edouard Belin, 31401 Toulouse Cedex 9, France\label{aff37}
\and
Space Science Data Center, Italian Space Agency, via del Politecnico snc, 00133 Roma, Italy\label{aff38}
\and
INAF-Osservatorio Astrofisico di Torino, Via Osservatorio 20, 10025 Pino Torinese (TO), Italy\label{aff39}
\and
Dipartimento di Fisica, Universit\`a di Genova, Via Dodecaneso 33, 16146, Genova, Italy\label{aff40}
\and
INFN-Sezione di Genova, Via Dodecaneso 33, 16146, Genova, Italy\label{aff41}
\and
Department of Physics "E. Pancini", University Federico II, Via Cinthia 6, 80126, Napoli, Italy\label{aff42}
\and
INFN section of Naples, Via Cinthia 6, 80126, Napoli, Italy\label{aff43}
\and
Instituto de Astrof\'isica e Ci\^encias do Espa\c{c}o, Universidade do Porto, CAUP, Rua das Estrelas, PT4150-762 Porto, Portugal\label{aff44}
\and
Faculdade de Ci\^encias da Universidade do Porto, Rua do Campo de Alegre, 4150-007 Porto, Portugal\label{aff45}
\and
Dipartimento di Fisica, Universit\`a degli Studi di Torino, Via P. Giuria 1, 10125 Torino, Italy\label{aff46}
\and
INFN-Sezione di Torino, Via P. Giuria 1, 10125 Torino, Italy\label{aff47}
\and
INAF-IASF Milano, Via Alfonso Corti 12, 20133 Milano, Italy\label{aff48}
\and
INFN-Sezione di Roma, Piazzale Aldo Moro, 2 - c/o Dipartimento di Fisica, Edificio G. Marconi, 00185 Roma, Italy\label{aff49}
\and
Centro de Investigaciones Energ\'eticas, Medioambientales y Tecnol\'ogicas (CIEMAT), Avenida Complutense 40, 28040 Madrid, Spain\label{aff50}
\and
Port d'Informaci\'{o} Cient\'{i}fica, Campus UAB, C. Albareda s/n, 08193 Bellaterra (Barcelona), Spain\label{aff51}
\and
Institute for Theoretical Particle Physics and Cosmology (TTK), RWTH Aachen University, 52056 Aachen, Germany\label{aff52}
\and
Institute of Cosmology and Gravitation, University of Portsmouth, Portsmouth PO1 3FX, UK\label{aff53}
\and
Dipartimento di Fisica e Astronomia "Augusto Righi" - Alma Mater Studiorum Universit\`a di Bologna, Viale Berti Pichat 6/2, 40127 Bologna, Italy\label{aff54}
\and
Jodrell Bank Centre for Astrophysics, Department of Physics and Astronomy, University of Manchester, Oxford Road, Manchester M13 9PL, UK\label{aff55}
\and
European Space Agency/ESRIN, Largo Galileo Galilei 1, 00044 Frascati, Roma, Italy\label{aff56}
\and
Universit\'e Claude Bernard Lyon 1, CNRS/IN2P3, IP2I Lyon, UMR 5822, Villeurbanne, F-69100, France\label{aff57}
\and
Institut de Ci\`{e}ncies del Cosmos (ICCUB), Universitat de Barcelona (IEEC-UB), Mart\'{i} i Franqu\`{e}s 1, 08028 Barcelona, Spain\label{aff58}
\and
Instituci\'o Catalana de Recerca i Estudis Avan\c{c}ats (ICREA), Passeig de Llu\'{\i}s Companys 23, 08010 Barcelona, Spain\label{aff59}
\and
UCB Lyon 1, CNRS/IN2P3, IUF, IP2I Lyon, 4 rue Enrico Fermi, 69622 Villeurbanne, France\label{aff60}
\and
Mullard Space Science Laboratory, University College London, Holmbury St Mary, Dorking, Surrey RH5 6NT, UK\label{aff61}
\and
Departamento de F\'isica, Faculdade de Ci\^encias, Universidade de Lisboa, Edif\'icio C8, Campo Grande, PT1749-016 Lisboa, Portugal\label{aff62}
\and
Instituto de Astrof\'isica e Ci\^encias do Espa\c{c}o, Faculdade de Ci\^encias, Universidade de Lisboa, Campo Grande, 1749-016 Lisboa, Portugal\label{aff63}
\and
Department of Astronomy, University of Geneva, ch. d'Ecogia 16, 1290 Versoix, Switzerland\label{aff64}
\and
INAF-Istituto di Astrofisica e Planetologia Spaziali, via del Fosso del Cavaliere, 100, 00100 Roma, Italy\label{aff65}
\and
INFN-Padova, Via Marzolo 8, 35131 Padova, Italy\label{aff66}
\and
School of Physics, HH Wills Physics Laboratory, University of Bristol, Tyndall Avenue, Bristol, BS8 1TL, UK\label{aff67}
\and
Universit\"ats-Sternwarte M\"unchen, Fakult\"at f\"ur Physik, Ludwig-Maximilians-Universit\"at M\"unchen, Scheinerstrasse 1, 81679 M\"unchen, Germany\label{aff68}
\and
FRACTAL S.L.N.E., calle Tulip\'an 2, Portal 13 1A, 28231, Las Rozas de Madrid, Spain\label{aff69}
\and
Dipartimento di Fisica "Aldo Pontremoli", Universit\`a degli Studi di Milano, Via Celoria 16, 20133 Milano, Italy\label{aff70}
\and
Institute of Theoretical Astrophysics, University of Oslo, P.O. Box 1029 Blindern, 0315 Oslo, Norway\label{aff71}
\and
Leiden Observatory, Leiden University, Einsteinweg 55, 2333 CC Leiden, The Netherlands\label{aff72}
\and
Jet Propulsion Laboratory, California Institute of Technology, 4800 Oak Grove Drive, Pasadena, CA, 91109, USA\label{aff73}
\and
Felix Hormuth Engineering, Goethestr. 17, 69181 Leimen, Germany\label{aff74}
\and
Technical University of Denmark, Elektrovej 327, 2800 Kgs. Lyngby, Denmark\label{aff75}
\and
Cosmic Dawn Center (DAWN), Denmark\label{aff76}
\and
Max-Planck-Institut f\"ur Astronomie, K\"onigstuhl 17, 69117 Heidelberg, Germany\label{aff77}
\and
NASA Goddard Space Flight Center, Greenbelt, MD 20771, USA\label{aff78}
\and
Department of Physics and Astronomy, University College London, Gower Street, London WC1E 6BT, UK\label{aff79}
\and
Department of Physics and Helsinki Institute of Physics, Gustaf H\"allstr\"omin katu 2, 00014 University of Helsinki, Finland\label{aff80}
\and
Aix-Marseille Universit\'e, CNRS/IN2P3, CPPM, Marseille, France\label{aff81}
\and
Universit\'e de Gen\`eve, D\'epartement de Physique Th\'eorique and Centre for Astroparticle Physics, 24 quai Ernest-Ansermet, CH-1211 Gen\`eve 4, Switzerland\label{aff82}
\and
Department of Physics, P.O. Box 64, 00014 University of Helsinki, Finland\label{aff83}
\and
Helsinki Institute of Physics, Gustaf H{\"a}llstr{\"o}min katu 2, University of Helsinki, Helsinki, Finland\label{aff84}
\and
European Space Agency/ESTEC, Keplerlaan 1, 2201 AZ Noordwijk, The Netherlands\label{aff85}
\and
Kapteyn Astronomical Institute, University of Groningen, PO Box 800, 9700 AV Groningen, The Netherlands\label{aff86}
\and
NOVA optical infrared instrumentation group at ASTRON, Oude Hoogeveensedijk 4, 7991PD, Dwingeloo, The Netherlands\label{aff87}
\and
Centre de Calcul de l'IN2P3/CNRS, 21 avenue Pierre de Coubertin 69627 Villeurbanne Cedex, France\label{aff88}
\and
INFN-Sezione di Milano, Via Celoria 16, 20133 Milano, Italy\label{aff89}
\and
University of Applied Sciences and Arts of Northwestern Switzerland, School of Computer Science, 5210 Windisch, Switzerland\label{aff90}
\and
Universit\"at Bonn, Argelander-Institut f\"ur Astronomie, Auf dem H\"ugel 71, 53121 Bonn, Germany\label{aff91}
\and
Dipartimento di Fisica e Astronomia "Augusto Righi" - Alma Mater Studiorum Universit\`a di Bologna, via Piero Gobetti 93/2, 40129 Bologna, Italy\label{aff92}
\and
Universit\'e Paris Cit\'e, CNRS, Astroparticule et Cosmologie, 75013 Paris, France\label{aff93}
\and
CNRS-UCB International Research Laboratory, Centre Pierre Binetruy, IRL2007, CPB-IN2P3, Berkeley, USA\label{aff94}
\and
University of Applied Sciences and Arts of Northwestern Switzerland, School of Engineering, 5210 Windisch, Switzerland\label{aff95}
\and
Institute of Physics, Laboratory of Astrophysics, Ecole Polytechnique F\'ed\'erale de Lausanne (EPFL), Observatoire de Sauverny, 1290 Versoix, Switzerland\label{aff96}
\and
Aurora Technology for European Space Agency (ESA), Camino bajo del Castillo, s/n, Urbanizacion Villafranca del Castillo, Villanueva de la Ca\~nada, 28692 Madrid, Spain\label{aff97}
\and
Institut de F\'{i}sica d'Altes Energies (IFAE), The Barcelona Institute of Science and Technology, Campus UAB, 08193 Bellaterra (Barcelona), Spain\label{aff98}
\and
School of Mathematics, Statistics and Physics, Newcastle University, Herschel Building, Newcastle-upon-Tyne, NE1 7RU, UK\label{aff99}
\and
DARK, Niels Bohr Institute, University of Copenhagen, Jagtvej 155, 2200 Copenhagen, Denmark\label{aff100}
\and
Perimeter Institute for Theoretical Physics, Waterloo, Ontario N2L 2Y5, Canada\label{aff101}
\and
Institute of Space Science, Str. Atomistilor, nr. 409 M\u{a}gurele, Ilfov, 077125, Romania\label{aff102}
\and
Consejo Superior de Investigaciones Cientificas, Calle Serrano 117, 28006 Madrid, Spain\label{aff103}
\and
Dipartimento di Fisica e Astronomia "G. Galilei", Universit\`a di Padova, Via Marzolo 8, 35131 Padova, Italy\label{aff104}
\and
Institut f\"ur Theoretische Physik, University of Heidelberg, Philosophenweg 16, 69120 Heidelberg, Germany\label{aff105}
\and
Institut de Recherche en Astrophysique et Plan\'etologie (IRAP), Universit\'e de Toulouse, CNRS, UPS, CNES, 14 Av. Edouard Belin, 31400 Toulouse, France\label{aff106}
\and
Universit\'e St Joseph; Faculty of Sciences, Beirut, Lebanon\label{aff107}
\and
Departamento de F\'isica, FCFM, Universidad de Chile, Blanco Encalada 2008, Santiago, Chile\label{aff108}
\and
Universit\"at Innsbruck, Institut f\"ur Astro- und Teilchenphysik, Technikerstr. 25/8, 6020 Innsbruck, Austria\label{aff109}
\and
Institut d'Estudis Espacials de Catalunya (IEEC),  Edifici RDIT, Campus UPC, 08860 Castelldefels, Barcelona, Spain\label{aff110}
\and
Satlantis, University Science Park, Sede Bld 48940, Leioa-Bilbao, Spain\label{aff111}
\and
Infrared Processing and Analysis Center, California Institute of Technology, Pasadena, CA 91125, USA\label{aff112}
\and
Instituto de Astrof\'isica e Ci\^encias do Espa\c{c}o, Faculdade de Ci\^encias, Universidade de Lisboa, Tapada da Ajuda, 1349-018 Lisboa, Portugal\label{aff113}
\and
Universidad Polit\'ecnica de Cartagena, Departamento de Electr\'onica y Tecnolog\'ia de Computadoras,  Plaza del Hospital 1, 30202 Cartagena, Spain\label{aff114}
\and
Centre for Information Technology, University of Groningen, P.O. Box 11044, 9700 CA Groningen, The Netherlands\label{aff115}
\and
INFN-Bologna, Via Irnerio 46, 40126 Bologna, Italy\label{aff116}
\and
INAF, Istituto di Radioastronomia, Via Piero Gobetti 101, 40129 Bologna, Italy\label{aff117}
\and
ICL, Junia, Universit\'e Catholique de Lille, LITL, 59000 Lille, France\label{aff118}}    

\abstract{
Intracluster light (ICL) provides a record of the dynamical interactions undergone by clusters, giving clues on cluster formation and evolution. Here, we analyse the properties of ICL in the massive cluster Abell~2390 at redshift $z=\,0.228$.
Our analysis is based on the deep images obtained by the \Euclid mission as part of the Early Release Observations in the near-infrared (\YE, \JE, \HE bands), using the NISP instrument in a 0.75\,deg$^2$ field. We subtracted a point--spread function (PSF) model and removed the Galactic cirrus contribution in each band after modelling it with the \texttt{DAWIS} software. We then applied three methods to detect, characterise, and model the ICL and the brightest cluster galaxy (BCG): the \texttt{CICLE} 2D multi-galaxy fitting; the \texttt{DAWIS} wavelet-based multiscale software; and a mask-based 1D profile fitting.
We detect ICL out to 600\,kpc. The ICL fractions derived by our three methods range between 18\,\% and 36\,\% (average of 24\,\%), while the BCG+ICL fractions are between 21\,\% and 41\,\% (average of 29\,\%), depending on the band and method. A galaxy density map based on 219 selected cluster members shows a strong cluster substructure to the south-east and a smaller feature to the north-west. 
Ellipticals dominate the cluster's central region, with a centroid offset from the BCG by about 70\,kpc and distribution following that of the ICL, while spirals do not trace the entire ICL but rather substructures. The comparison of the BCG+ICL, mass from gravitational lensing, and X-ray maps show that the BCG+ICL is the best tracer of substructures in the cluster.
Based on colours, the ICL (out to about 400~kpc) seems to be built by the accretion of small systems ($M \sim 10^{9.5}\si{\solarmass}$), or from stars coming from the outskirts of Milky Way-type galaxies ($M \sim 10^{10}\si{\solarmass}$). Though Abell~2390 does not seem to be undergoing a merger, it is not yet fully relaxed, since it has accreted two groups that have not fully merged with the cluster core. We estimate that the contributions to the inner 300\,kpc of the ICL of the north-west and south-east subgroups are 21\,\% and 9\,\% respectively.
}
% Provide up to five key words:
%
%    from the list in
%     https://www.aanda.org/for-authors/latex-issues/information-files#pop}
%
% Add short versions of title and author list for page headings
%
   \titlerunning{\Euclid: ERO -- The intracluster light of A\,2390}
   \authorrunning{Ellien et al.}
   
   \maketitle
%
%-------------------------------------------------------------------
%
%
%   Start the main text of your paper here
%

\section{\label{sc:Intro}Introduction}
As galaxies in groups and clusters interact, stars are ejected from their galactic moorings and end up populating the space between the galaxies. Over time, these unbound stars form the intracluster light (ICL), a characteristic diffuse glow seen throughout groups and clusters see \cite{Contini2021} and 
\citet[][]{Montes2022} for reviews. As a by-product of the interactions between the cluster galaxies, the ICL is a fossil record of all the dynamical interactions that the system has experienced \citep[e.g.,][]{Merritt1984, Gregg1998}. The ICL therefore provides a holistic view of the history of the cluster. As such, the formation and assembly history of the ICL is central to understanding the global evolution of galaxy clusters.

The stellar populations of the ICL reflect the properties of the galaxies from which it has accreted its stars. Therefore, studying this light allows us to infer the mechanisms involved in forming this component. Simulations have suggested several mechanisms that could be responsible for the formation of the ICL: total disruption of low-mass satellites \citep[][]{Purcell2007, Barai2009}; tidal stripping of massive satellites \citep[e.g.,][]{Rudick2009, Contini2014, Contini2019}; stars ejected into the intracluster medium after a merger \citep[][]{Willman2004, Murante2007, Conroy2007}; in-situ star formation \citep{Puchwein2010}; and accretion of the ICL from groups \citep[`pre-processing',][]{Mihos2004}. Each mechanism leaves a distinct imprint on the properties of the stellar populations of the ICL. 

Over the last 20 years, observations have shown that the ICL is a ubiquitous feature of clusters \citep[e.g.,][]{Feldmeier2004, Kluge2020, golden-marx2023, Ragusa2023, golden-marx2024}. However, the ICL is extended and faint 
\citep[$\mu_v >$26.5\,mag\,arcsec$^{-2}$,][]{Rudick2006}, making it challenging to obtain good-quality observational data. Consequently, for most systems, we only have access to broad-band imaging. This is even more difficult in the infrared (IR), where the brightness of the Earth's atmosphere 
\citep[13--14\,mag\,arcsec$^{-2}$, ][]{Oliva2003}\footnote{\url{https://www.eso.org/gen-fac/pubs/astclim/espas/espas_reports/ESPAS-MaunaKea.pdf}} rivals that of low surface brightness (LSB) features such as the ICL. As a result, observations of the ICL and other LSB features have largely been limited to the blue part of the spectrum. Space-based observations give us the opportunity to explore the LSB Universe in the IR. 

The \Euclid \citep{Laureijs11,EuclidSkyOverview} space mission will observe nearly one-third of the sky in four photometric wavebands: the visible band (\IE) using the VIS instrument \citep{EuclidSkyVIS}; and three near-infrared (NIR; \YE, \JE, \HE) bands using the NISP instrument \citep{EuclidSkyNISP}. \Euclid's faint detection limit and wide field-of-view (FoV) make it an ideal instrument for studying the LSB Universe, particularly the diffuse ICL, across a large redshift range \citep{Scaramella-EP1,Borlaff-EP16}. Moreover, including IR wavelengths in studying the stellar populations will better constrain their properties (age and metallicity) than optical data alone \citep{Worthey1994}.

This work focuses on the intermediate redshift cluster Abell 2390 (A\,2390 hereafter), a well-known cool-core cluster, with a brightest cluster galaxy (BCG) located at \mbox{${\rm RA}=\ra{21;53;36.83}$}, \mbox{${\rm Dec}=\ang{+17;41;43.73}$} (all analysis used in this paper is centred on this BCG) with a redshift $z=0.228$ \citep[e.g.,][]{Abraham1996, sohn20}. The images used in this analysis were taken as part of the Early Release Observations (ERO) programme, a series of observations that illustrate \Euclid's capabilities \citep[][]{EROcite}. 
A\,2390 is a massive cluster, with {$M_{200,\rm{c}}$} between $1.53 \times 10^{15}$ \si{\solarmass} \citep[weak lensing;][]{okabe16} and $1.84 \times 10^{15}$$ \si{\solarmass}$ \citep[projected phase-space of galaxies;][]{sohn20} and a virial radius of $R_{200, \rm{c}} = 2.1\,{\rm Mpc}$ \citep{Carlberg1997}. Prior works have characterised some of A\,2390's formation history.  For example, \citet{Abraham1996} found that the cluster was built up gradually by the infall of field galaxies over around $8\,{\rm Gyr}$. Moreover, X-ray data also reveal the presence of a cooling flow associated with the BCG and the presence of an active galactic nucleus \citep[AGN;][]{Allen2001, Alcorn2023}. Additionally, radio observations show the presence of an extended double lobe located $300\,{\rm kpc}$ east and west of the BCG, remnants of past AGN activity \citep{Savini2019, Alcorn2023}.

The paper is structured as follows: we briefly describe in Sect.~\ref{sec:observations} the data used in this work, in Sect.~\ref{sec:data_processing} the processing and cleaning steps performed on the \Euclid images before ICL analysis, in Sect.~\ref{sec:Method} the methods used to detect the ICL, and in Sect.~\ref{sc:Results} the results of these methods and the comparison of ICL properties with other cluster components. Finally, we discuss in Sect.~\ref{sec:discussion} the implications of these results and give our conclusions in Sect.~\ref{sec:conclusions}.

Throughout this analysis, we assume that A\,2390 has a redshift of $z=0.228$ \citep{sohn20}. We also assume a standard flat Lambda cold-dark-matter ($\Lambda$\,CDM) cosmology with $\Omega_{\rm m}=0.3$ and $H_0=70$\,km\,s$^{‐1}${\rm Mpc}$^{‐1}$ \citep{Planck_2018}. Lastly, all magnitudes presented in this work are in the AB system.

\section{Observations}
\label{sec:observations}

The data used in this analysis were taken as part of the \Euclid ERO programme \cite[][]{EROcite}, which targetted the lensing clusters of galaxies A\,2390 and Abell 2764, as illustrated in \citet{EROData} and \citet{EROLensdata}, where the observational approach and photometric data reduction methodology used in this analysis are thoroughly described. 
Here we summarise the relevant portions of those analyses. As part of the ERO, A\,2390 was observed for three reference observing sequences, which is three times the \Euclid Wide Survey (EWS) exposure time, allowing us to potentially probe the ICL to fainter magnitudes and larger radial extents than we anticipate in the EWS (Bellhouse et al. in prep.).  Since this paper focuses on the faint ICL component we only use the data reduction method that preserves the LSB features.

The FoV of the observations is 0.75\,deg$^{2}$. This pointing is centred on A\,2390's cluster core, which represents a small fraction of the total area ($R_{200, {\rm c}}$ = 0\fdg16). The \IE\ images have a pixel scale of 0\,\arcsecf 1\, and a spatial resolution (FWHM) of \ang{;;0.16}. The NISP images have a pixel scale of 0\,\arcsecf 3\,pix$^{-1}$ and a spatial resolution of $\ang{;;0.49}$. For reasons described in Sect. \ref{sec:cirri}, we only use the NISP images in this work. Additionally, we note that the surface brightness limits ($3\,\sigma$, $10\,\arcsec\times10\,\arcsec$) of the NISP images are 28.7, 28.9, and 29.0 mag arcsec$^{-2}$ for the \YE, \JE, and \HE bands, respectively, following appendix A in \citet{Roman2020}.

\section{Data processing}
\label{sec:data_processing}

\subsection{Point spread function}
\label{sec:psf}

We use for this analysis the well-sampled and modelled PSF derived in \citet{EROData} for each NISP image (\YE, \JE, and \HE). As was the case in \citet{Kluge2024}, our goal is not to accurately subtract the PSF core from the saturated bright stars, but instead to remove the star's outer light profile using this PSF model, since the outer profile includes light that can be mistakenly identified as ICL \citep[e.g.,][]{Montes2021}. 

For the near-infrared (NIR) bands, the PSF used in this analysis was 350 pixels ($105$\,\arcsec) in radius. The PSF model is then subtracted from the 70 brightest stars in the field of A\,2390. We exclude stars near the FoV's edge or close to another bright source (which prevented the star's outer profile from being properly subtracted). We subtracted the same stars in the \YE, \JE, and \HE\ images. To optimise the subtraction of the outer profiles, we select the aperture around the star used to normalise the PSF profiles by identifying the aperture that minimises the $\chi^2$ statistic for the difference between the normalised PSF and each star's outer light profile.

\begin{figure}
\begin{center}
\includegraphics[width=\columnwidth]{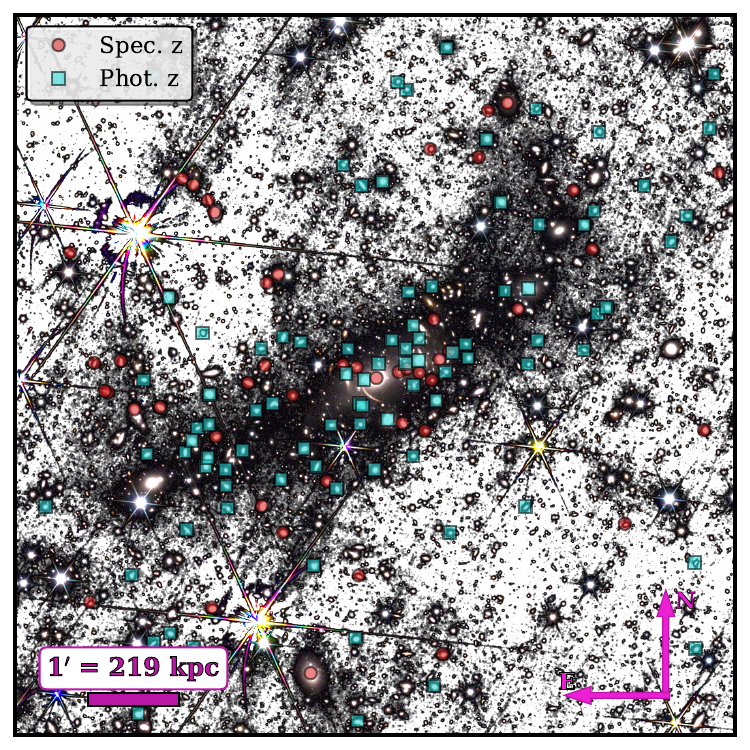}
\end{center}
\caption{Image of the $\ang{;9.1;}\times \ang{;9.1;}\, (2$ Mpc $\times\, 2$ Mpc, approximately 0.5$R_{200}$ of the cluster) region around A\,2390. The image is a combination of an RGB image of the PSF-subtracted NISP images and a $\JE+ \HE$ black and white background. Cluster members are colour-coded according to whether their redshift is spectroscopic (red circles) or photometric (teal squares). North is up and east is to the left. This image highlights the importance of removing bright stars close to the cluster centre to study the diffuse light (see Sect.~\ref{sec:psf}). }
\label{fig:rgb_a2390}
\end{figure}

\subsection{Galactic cirrus modelling and subtraction}\label{sec:cirri}

In the context of LSB astronomy, the so-called Galactic cirrus is a reflection-induced signature of the Galaxy's interstellar medium (ISM) in optical and NIR images, which is present up to high Galactic latitudes \citep{Planck2016} due to its proximity. It displays a complex filamentary-like pattern that can mimic the shape and brightness of faint extragalactic features \citep{Duc2015}.

\begin{figure}
\begin{center}
\includegraphics[width=\columnwidth]{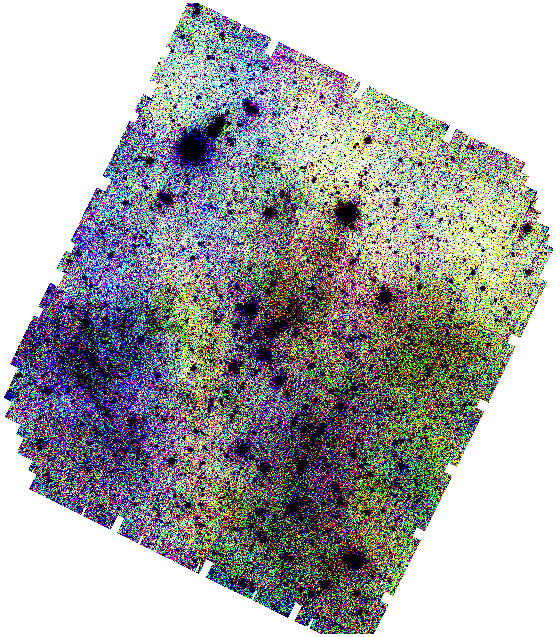}
\end{center}
\caption{RGB image of the A\,2390 galaxy cluster \Euclid\ FoV. The contrast is greatly enhanced to highlight large-scale colour variations (due to cirrus and background inhomogeneities) across the whole image in the LSB regime. Blue is VIS, green is the \JE band, and red is the \YE band. North is up and east is to the left.}  
\label{fig:cirrus_rgb_VIS_Y_J}
\end{figure}

With the increasing depth of the new observations, Galactic cirrus has become a pervasive component of the optical images, often occupying a large fraction of the FoV. It is a source of light contamination to the ICL \citep{Mihos2017} as both components are mixed in a non-trivial way, and deblending them is a challenge. This is the case for A\,2390; the images (as presented in Fig.~\ref{fig:cirrus_rgb_VIS_Y_J}) show a great amount of foreground cirrus needing to be accounted for before the ICL can be measured. 

Previous papers have studied the wavelength dependency of Galactic cirrus, showing that the dust-scattered component is prominent in optical bands with a decreasing albedo in the NIR \citep{Roman2020, Zhang2023}. The trend is similar in A\,2390 images, as more prominent cirrus features were observed in the VIS image compared to the NISP ones. The least affected band seems to be \HE. 
In the case of A\,2390, the cirrus presents colour variations across the FoV, especially comparing VIS to NISP (see Fig.~\ref{fig:cirrus_rgb_VIS_Y_J}), indicating a heterogeneous ISM with a spatially and spectrally varying composition, probably due to the varying column density of the dust clouds \citep[e.g.,][]{Roman2020}. 

\subsubsection{FIR dust maps}\label{sec:wise}
%{Choice of methodology}

At longer wavelengths, cirrus has its peak emission in the FIR due to thermal emission from low-temperature dust \citep{Low1984, Veneziani2010} which roughly correlates with optical surface brightness \citep{Witt2008}. This led previous ICL studies to assume a spatial match between FIR dust maps and optical/NIR cirrus light \citep{Mihos2017, Kluge2020}. In particular, \citet{Kluge2024} corrected for significant cirrus contamination in \Euclid's ERO VIS image of the Perseus cluster by empirically scaling the \textit{Planck}/WISE 12-\micron\ dust map.

A procedure similar to \citet{Kluge2024} is attempted to remove the foreground cirrus in the A\,2390 images, using the dust emission maps from \citet[][]{Meisner2014}. This map is generated from the WISE 12-\micron\ imaging data and is free of compact sources and other contaminating artefacts. The angular resolution of these dust maps is 26\arcsec\ per pixel. We tried different normalizations of this dust map to match the average background properties of the \Euclid\ images. Unfortunately, none of the normalizations provides a reliable cirrus subtraction, leaving a significant amount of filamentary cirrus in the residuals, especially over the location of A\,2390. An example is shown in Appendix \ref{app:cirri}. In this case, the resolution of the dust maps does not describe the distribution of the cirrus. Since the cluster's size is similar to the size of the dust filaments, this method is not optimal. 

The different spatial resolutions between the FIR maps and optical observations are the main source of this disparity. In addition, the FIR to optical/NIR similarity is a first-order approximation to the complex multi-phase Milky Way (MW) ISM, which may include different populations of dust grains (composition, size, and geometry) in a variety of physical conditions (local radiation field, density, and temperature), see \cite{Draine2007} and \cite{Brandt2012}.

Following these preliminary tests, we concentrate on the NISP images that are less contaminated by cirrus.
As the cirrus covers the whole FoV, most of its distribution varies over angular scales much larger than the extent of A\,2390.
Therefore, an alternative empirical wavelet-based approach is chosen for direct cirrus modelling and removal in the NISP images (see next section).

\subsubsection{Multiscale modelling}
\label{subsubsec:multiscale_modeling}

\begin{figure*}
\begin{center}
\includegraphics[width=\textwidth]{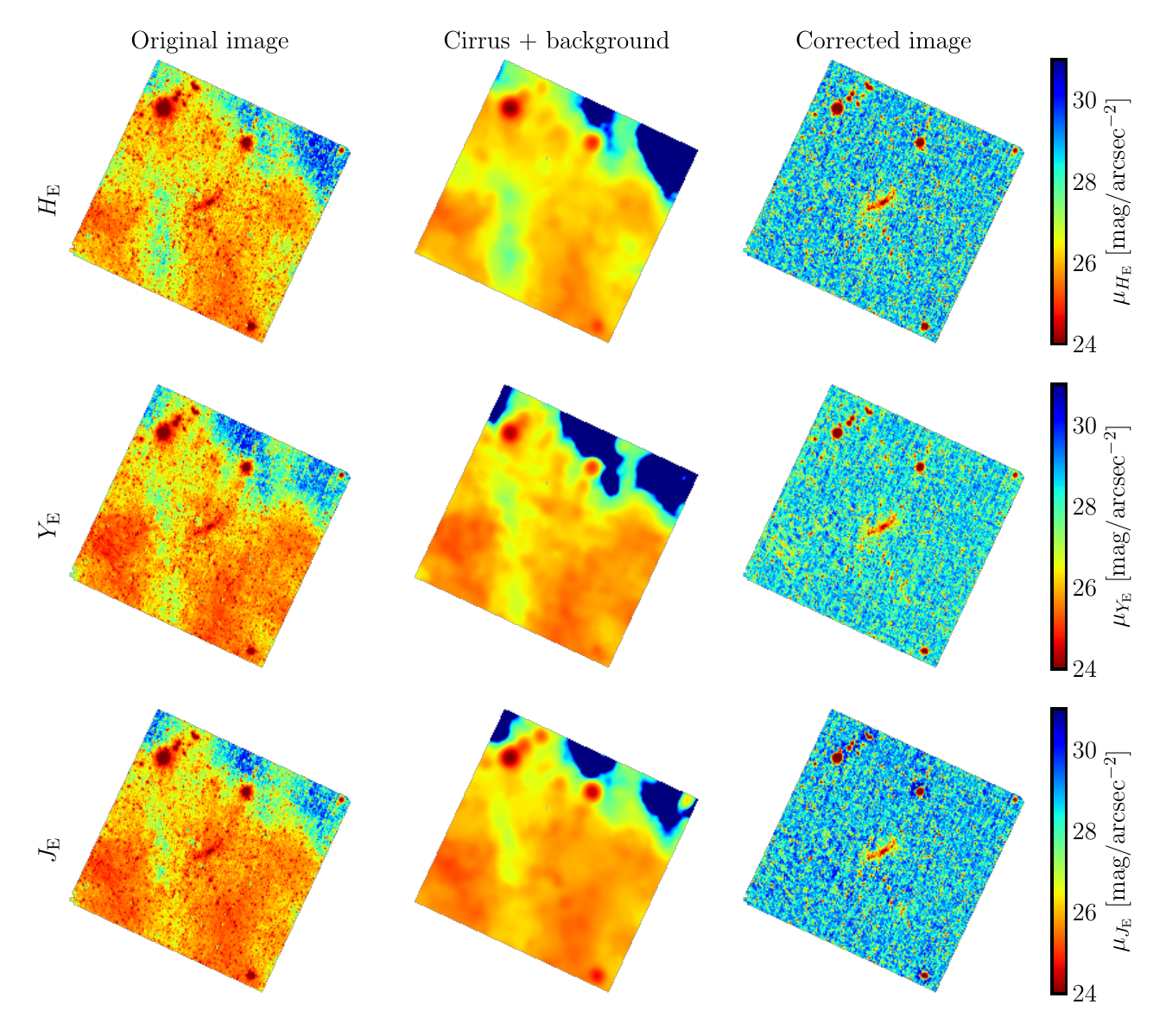}
\end{center}
\caption{Effect of cirrus removal for all NISP filter images. From left to right: original image; cirrus+background map; cirrus-corrected image. The angular size of the images is $0\degf 7\times 0\degf 7$.}
\label{fig:cirrus_clean}
\end{figure*}

\begin{figure*}
\begin{center}
\includegraphics[width=\textwidth]{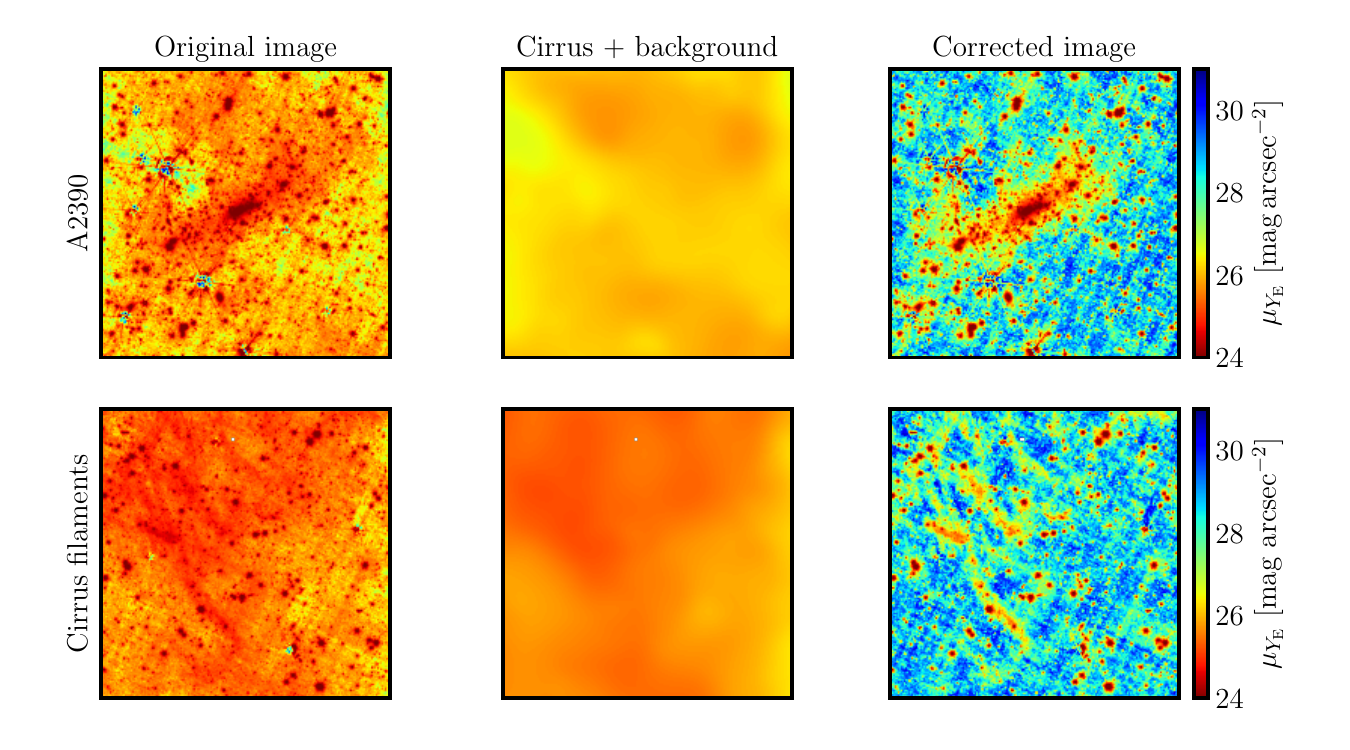}
\end{center}
\caption{Zoom-in on regions of interest in the \YE band, which displays the most cirrus residuals after correction. The top row shows a $\ang{;12.5;}\,\times\,\ang{;12.5;}$ image centred on 
A\,2390. The bottom row shows a high-intensity cirrus region ($\ang{;12.5;}\,\times\,\ang{;12.5;}$) situated in the east of the image, displaying obvious small-scale cirrus residuals in the corrected image. From left to right: original image; cirrus model; cirrus-corrected image.}
\label{fig:cirrus_residuals}
\end{figure*}

We perform a multiscale decomposition of the images to disentangle the large-scale cirrus signal (covering most of the FoV) from the smaller angular scale (1 to a few arcmin) extragalactic sources, ICL included.
This part of the analysis is based primarily on the Detection Algorithm with Wavelets for Intracluster light studies \citep[\texttt{DAWIS};][]{Ellien2021}. The detailed operating process of \texttt{DAWIS} is described extensively in \cite{Ellien2021}, so only salient points are summarised here. \texttt{DAWIS} leverages a wavelet representation \citep{Slezak1994} and multi-resolution vision models \citep{bijaoui1995} to separate small-scale details from large-scale variations in the analysed images. For this purpose an isotropic undecimated wavelet transform \citep{Starck2007} is applied to the image, decomposing it into $N$ wavelet planes of the same size. The noise is estimated and modelled in the first wavelet scale before being extrapolated to larger scales to detect sources by thresholding these maps. These regions of significant wavelet coefficients are linked into interscale trees, which are then used to reconstruct the corresponding two-dimensional light distribution in the original image. The distinctive feature of \texttt{DAWIS} is its iterative strategy: it models only a few sources at a time (controlled by a threshold factor $\tau$) starting with the brightest and removing a fraction of their 2D light profile (controlled by a mitigation parameter $\gamma$) from the image. The whole process (from the wavelet transform to light profile modelling) is repeated until it converges on a residual map that contains only noise. At each iteration, the detected and modelled sources correspond to substructures rather than entire astrophysical objects. These are termed `atoms'; since the sum of all these light contributions reproduces a fully noise-free version of the astrophysical field. One can also select a fraction of the atoms based on criteria (size, morphology, wavelet scale...), and sum their light profiles to synthesise and model specific astrophysical sources.

For this work, \texttt{DAWIS} is complemented by a mixture of \texttt{gnuastro} \citep{Akhlaghi2015} and \texttt{photutils} \citep{Bradley2024} routines. First, the NISP images are pre-processed to be suitable for a wavelet analysis: the \texttt{astwarp} routine is used to rotate the images ($\Theta=65\,\degree$) so the two axes of the image correspond to the two axes of the wavelet transform. The images are then cropped with \texttt{astcrop} to a box of size $0\degf 7\times 0\degf 7$ covering all but a small fraction of the FoV. To reduce the computing time of \texttt{DAWIS}, the cropped images are binned by a factor of 4, resulting in images of size $2251\,{\rm pix}\times2251\,{\rm pix}$ with pixel scale 1\arcsecf2. 
As the goal is to model and remove the large-scale cirrus background, the loss of some small-scale information inserted into the analysis by these operations is not an issue. Moving forward, \texttt{DAWIS} is run on the resulting images with input parameters set to ensure a quick but rough-quality reconstruction of small and compact sources. This ensures that the algorithm quickly reaches the larger wavelet scales. Therefore, following values advocated by \citet{Ellien2021}, $\tau$ and $\gamma$ are set to 0.1 and 1 respectively.

The two assumptions made to select the cirrus distribution atoms are: i) its distribution varies over larger angular scales than the ICL, so its information is encoded in atoms detected at lower-frequency wavelet scales, ii) its distribution should not feature a peak centred on A\,2390 position on the sky, as this would rather be ICL.
After visual inspection of the outputs, the cirrus maps are derived by selecting atoms detected at wavelet scales $\geq\,6$, ensuring that we only keep sources larger than a few arcmin. 
Note here that some bright foreground MW stars that are not PSF-subtracted in the images are larger than this characteristic size, so part of their atoms are also selected and included in the cirrus maps. 
This is not a problem in this analysis because these stars are located far from the full extent of A\,2390.
To minimise removal of ICL from A\,2390, a hand-made ellipse ($\ang{;0.5;}\,\times\,\,2\,\arcmin$ with a $30\degree$ angle and centred on \mbox{${\rm RA}=\ra{21;53;38.03}$}, \mbox{${\rm Dec}=\ang{17;41;40.44}$}) is made to roughly cover the cluster extent. 
All atoms with peak coordinates of their light distributions within the ellipse are not included in the cirrus map.
Additional attention is brought to the \texttt{DAWIS} residuals, and a large-scale gradient is still noticeable after the whole wavelet procedure. Since the cirrus signal occupies most of the FoV, it is difficult to distinguish it from the true sky background, so both are combined into a cirrus + background map.
To do so, the residuals are fitted with the \texttt{Background2D} function from \texttt{photutils}, with a box size of $40\,{\rm pix}\times40\,{\rm pix}$ and a median filter of size $3\,{\rm pix}\times3\,{\rm pix}$, and the resulting 2D background is added to the cirrus map.
The cirrus maps are then scaled back to the original NISP pixel scale and orientation. The cirrus-corrected images are then produced by subtracting these from the original image, as shown in Fig.~\ref{fig:cirrus_clean}. A first-order estimation of the surface brightness cirrus model is made by using as sky background the mean value of the noise in a vertical strip of width $\ang{;0.1;}$\, on the right edge of the original images (which show less cirrus contamination, see Fig.~\ref{fig:cirrus_clean}). This gives median surface brightness values for the cirrus model of 26.3, 26.1, and 26.1 mag arcsec$^{-2}$, with maximum peaks of 23.9, 24.2, and 23.9 mag arcsec$^{-2}$ for the brightest filaments in the \YE, \JE, and \HE bands respectively.\footnote{These values do not account for zodiacal light contribution as we do not try to isolate cirrus from it.} The surface brightness limits ($3\sigma$, $10\,\arcsec\times\,10\,\arcsec$) of the cirrus-subtracted NISP images are 29.3, 29.4, and 29.4 mag arcsec$^{-2}$ for the \YE, \JE, and \HE bands, respectively, which is about 0.4 mag deeper than the original images.

\subsubsection{Limitations and uncertainties}
\label{subsec:limitations_and_uncertainties}
An intrinsic issue to the empirical process described in the previous section is the inability of an artificial wavelet scale separation to capture all subtleties of the hierarchical cirrus light distribution. While most of the large-scale cirrus is captured and modelled, some of the finer filaments (e.g., of similar size as extragalactic sources) are visible in the cirrus-corrected images (especially in the \YE band, see Fig.~\ref{fig:cirrus_residuals}). This inevitably leads to some flux-positive cirrus contamination left in the corrected images that might locally increase the ICL light level and bias colour measurements. For similar reasons, another effect occurs as some of the larger scale ICL can be accidentally included in the cirrus map and removed from the image. To estimate the uncertainties resulting from these effects and their influence on the ICL, a series of tests is performed, based on mock clusters inserted in the \HE image.

The process of creating the mock clusters and their ICL is similar to the one described in depth in Bellhouse et al. (2024, in prep.), so we only provide a quick summary here. The most massive cluster from the MAMBO simulated light-cone catalogue \citep{girelli2021} is used for the mocks.
The ICL, BCG, and satellite galaxy images are produced using the \texttt{galsim} \citep{Rowe2015} package. Satellite galaxies are modelled with S\'ersic profiles having either single-component or two-component bulge+disk profiles, whilst the BCG and ICL are generated as S\'ersic components using the mean values of double-S\'ersic decompositions performed by \citet{Kluge2020}, scaled to the stellar mass of the cluster. All light profiles are convolved with the \Euclid PSF.
The same cluster is inserted with the same orientation at nine different positions within the \Euclid observation chosen to cover a varied range of cirrus intensities to measure the effect of subtracting different levels of cirrus on the resulting flux of the ICL.

\begin{figure}
\begin{center}
\includegraphics[width=\columnwidth]{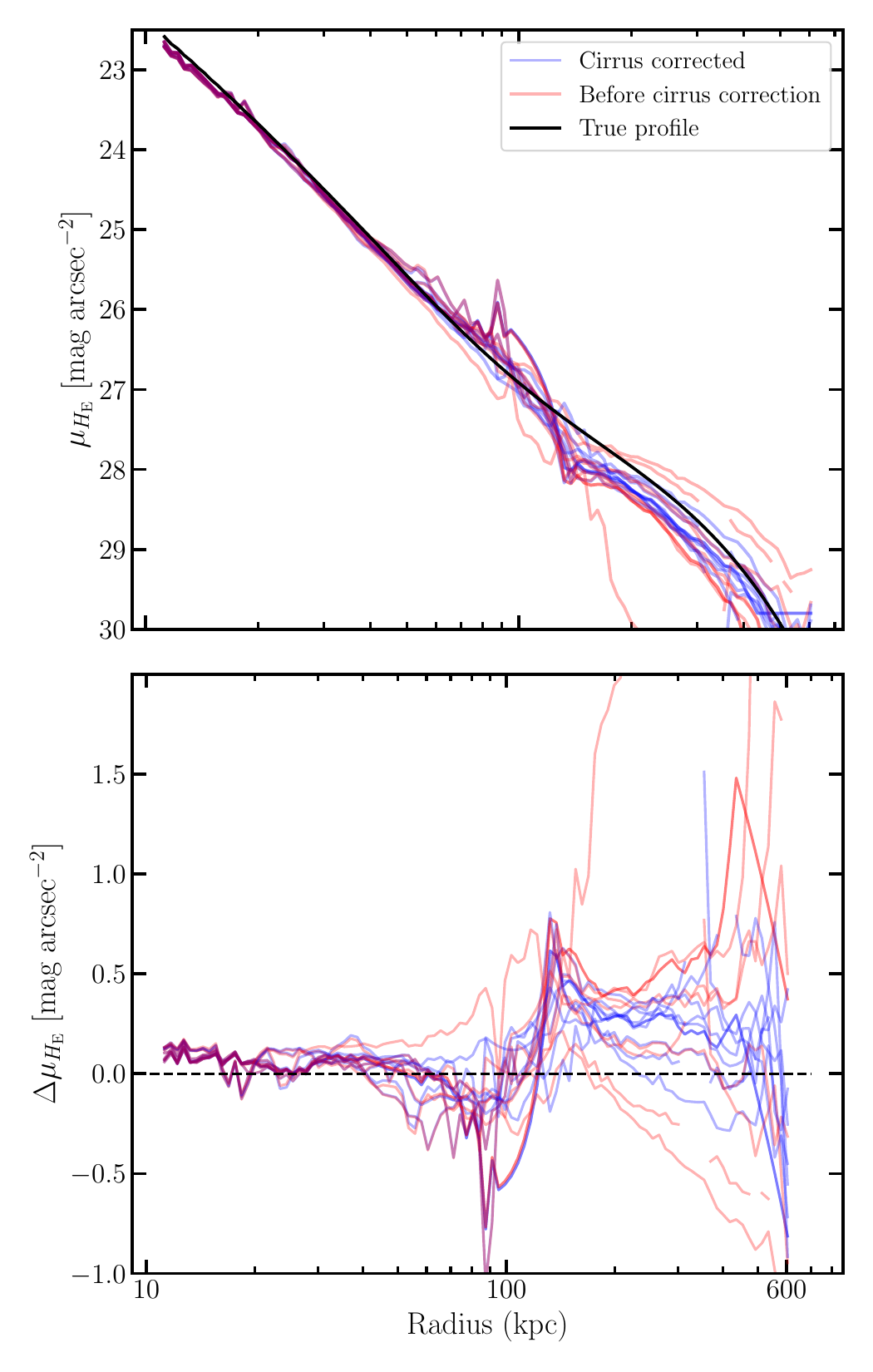}
\end{center}
\caption{Top: mock cluster radial profiles in the \HE band before (red) and after (blue) cirrus correction. The true BCG+ICL light profile is shown as a black line. Bottom: difference between true BCG+ICL light profile and the measured BCG+ICL profiles.}
\label{fig:cirrus_test_mock_profiles}
\end{figure}

The same cirrus modelling process as in Sect.~\ref{subsubsec:multiscale_modeling} is applied to the image with the mock clusters. BCG+ICL radial intensity profiles are derived in circular aperture for all mocks, before (red) and after (blue) the cirrus correction. Figure~\ref{fig:cirrus_test_mock_profiles} shows the BCG+ICL profiles before and after cirrus correction, alongside the true BCG+ICL profile used for the mock clusters and the difference between them. The effect of the cirrus on the profiles is the strongest in the outer part (>\,100~kpc), as the profiles before cirrus correction display significant disparity depending on local cirrus properties and large deviations from the true values (reaching values larger than 2 orders of magnitude for most of the clusters). After cirrus correction this disparity is greatly reduced, reducing the deviation from the true profile to absolute values below 0.5 orders of magnitude for most clusters, out to 600\,kpc and down to surface brightness values of $30$ (see lower panel of Fig.~\ref{fig:cirrus_test_mock_profiles}).

\subsection{Selection of cluster member galaxies}
\label{sec:members}

We extract a catalogue of all galaxies with redshifts available within a radial aperture of 30\,\arcmin\, around A\,2390 from the NASA Extragalactic Database (NED).\footnote{\hyperlink{https://ned.ipac.caltech.edu/}{https://ned.ipac.caltech.edu/}} There are 488 galaxies with spectroscopic redshifts. The redshift histogram peaks at the cluster redshift, and is quite clearly limited to the $\brackets{0.2169,0.2369}$ interval, which we consider hereafter as the cluster range. There are 184 galaxies with redshifts in this interval. The corresponding velocity range is $\brackets{58\,147, 62\,839}$~\SI{}{\kilo\meter\per\second}.

In addition to the spectroscopic members, we supplement our membership (particularly in the core of A\,2390) using the high-probability members ($P_{\rm mem}$ > 0.8) from the SDSS-redMaPPer \citep{rykoff2014} catalogue.  We note that redMaPPer only uses galaxies with SDSS $i < 21.0$ and that the red-sequence-based cluster redshifts provided by redMaPPer are consistent with existing spectroscopic redshifts \citep{rykoff2014}.  28 of these redMaPPer galaxies are spectroscopic members (matched members within a 0\arcsecf1
radial distance of our spectroscopic member catalogue described above) and the SDSS photometry provides us with more accurate galaxy positions. After visual inspection of each source, this results in a catalogue of 219 galaxies that can be considered as belonging to the cluster. This catalogue is used to compute a galaxy density map of cluster members (see Sect.~\ref{subsec:galaxy_density_map}), as well as galaxy morphology distributions (see Sect.~\ref{subsec:morph}).

The cluster member catalogue is then cross-matched with the ERO object catalogue \citep{EROData}, which contains photometric information for the four \Euclid bands of the whole FoV, obtained by running \citep[\textsc{SExtractor};][]{Bertin1996} on both the VIS and NISP data. The MW extinction corrections for each Euclid filter are derived using the Planck thermal dust map \citep{Planck_dust2014, Gordon2023} extinction law, assuming a $5700$\,K blackbody spectral energy distribution. 

\subsection{Masks}
\label{sec:mask_1d}
To accurately measure the properties of the diffuse light of 
A\,2390, all sources except the BCG and ICL must be masked. We use the PSF-cirrus-subtracted NISP  images in the separate \YE, \JE, and \HE bands from Sect.~\ref{subsubsec:multiscale_modeling} to start the masking process. Next, the masks are combined to obtain a final NISP mask ($\YE+\JE+\HE$). Additionally, while subtracting the stellar PSF model effectively removes the diffuse extended wings of the PSF (Sect.\,\ref{sec:psf}), the central regions show significant residuals. Therefore, the bright star central regions are masked with a circular patch of 70 pixels (210\,\arcsec) in radius.

As these are deep images, the masking must be optimised for faint and compact background objects as well as those that are larger. To do so, we use \textsc{SExtractor} in a `hot+cold' masking mode, similar to the method used in \citet[][]{MT14,MT18} and \cite{Montes2021}. The `hot' mode is optimised to detect the small and faint sources and is run using the following steps: 
\begin{enumerate}[i)]
    \item the contrast of the image is enhanced by creating an unsharp-masked version in each band;
    \item the unsharp-masked image is made by convolving the original image with a box filter with a side of 15 pixels (45\,\arcsec) and then subtracting the convolved image from the original;
    \item \textsc{SExtractor} is run on the unsharp-masked image with the source detection threshold of $1\sigma$ above the background.
\end{enumerate}
The `cold' mode is used to find large and diffuse sources, which is achieved by running \textsc{SExtractor} with a minimum source size of 40 pixels and a $5\sigma$ detection threshold. Finally, the hot and cold modes are combined along with the circular bright star masks to create three different masks: (i) a mask for all sources; (ii) a mask for all sources except the BCG+ICL, and (iii) a mask for all sources but the BCG+ICL and cluster members (see Sect.~\ref{sec:members} for details). 
 
Each mask requires some additional steps. The mask for all sources is created to measure any potential residual background light in the image (see Sect. \ref{sec:bkg_profiles_1d}). Therefore, it is essential to mask the ICL along with the galaxies for this step. To achieve this, the cold masks are radially extended by 15\,kpc (4\arcsec) and the hot masks are radially extended by 5\,kpc (1\farcs4) at the cluster redshift. These extensions are chosen after visual inspection. Finally, a $3\sigma$ clipping is applied around the mean value of the remaining pixels to mask any residual high/low-value pixels not detected by \textsc{SExtractor} during the masking process above. This last sigma clipping step is not done in masks (ii) and (iii) because this process also masks some pixels from the extended BCG light. Masks (ii) and (iii) are also prepared for a $2\,{\rm Mpc}\times2\,{\rm Mpc}$ cutout around the cluster centre, while mask (i) is made for the entire A\,2390 \Euclid FoV. 

\section{Measuring the ICL of A\,2390}
\label{sec:Method}
%\subsection{Detection Methods}

Although the qualitative definition of ICL as light emitted by stars not bound to the galaxy, distributed throughout the cluster's gravitational potential, is straightforward, its photometric observational signature is ambiguous, especially regarding its strong and smooth entanglement with the BCG light profile. This lack of consensus stacks on top of other generic LSB astronomy challenges (such as cirrus contamination, see Sect.\,~\ref{sec:cirri}) making the detection and characterisation of ICL challenging. As a result, many ICL measurement strategies have been developed by observers, often independently from one another. This lack of a definition inevitably leads to disparities when comparing results such as ICL and BCG+ICL fractions. Recently, \citet{Brough2024} addressed this issue by testing several observational methods on state-of-the-art simulations, finding that the different methods are consistent.

Inspired by the robust multiplex analysis presented in \citet{Brough2024}, three of the methods used in that analysis are used here to detect, characterise, and model the ICL: 2D multi-galaxy fitting \citep[\texttt{CICLE}, ][]{jimenez-teja2016}; wavelet-based multiscale analysis \citep[\texttt{DAWIS}, ][]{Ellien2021}; and mask-based 1D profile fitting \citep{Ahad2023}. These methods are described in more detail in the following sub-sections.

\subsection{CICLE}\label{sec:cicle}

\begin{figure}
\begin{center}
\includegraphics[width=\columnwidth]{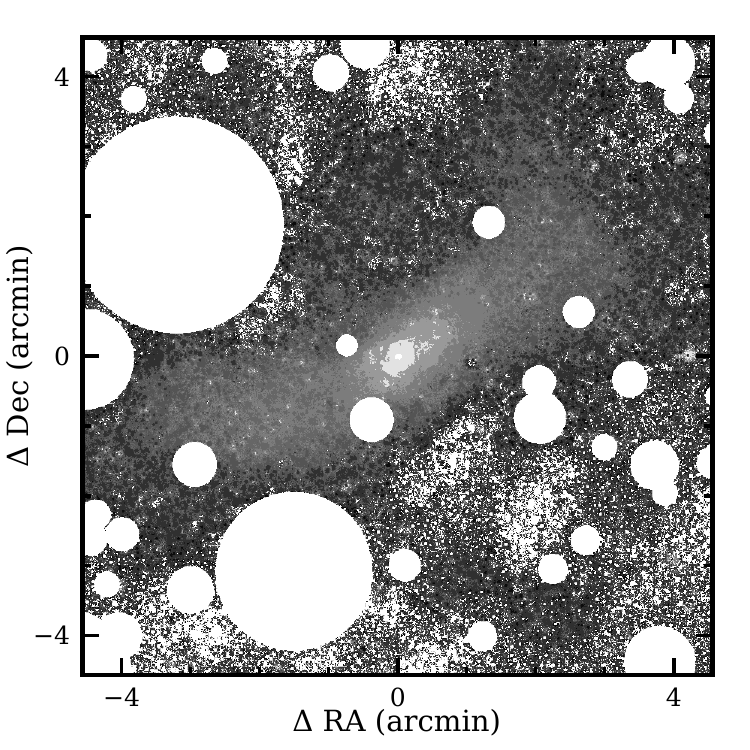}
\end{center}
\caption{ A $\ang{;9.1;}\times\ang{;9.1;}$ ($2\,{\rm Mpc}\times2\,{\rm Mpc}$) region around A\,2390 on the \texttt{CICLE} BCG+ICL \JE map. The white circles are masks over stars in the FoV. }  
\label{fig:cicle_j}
\end{figure}

The CHEFs Intracluster Light Estimator \citep[\texttt{CICLE}, ][]{jimenez-teja2016} is a multi-galaxy fitting algorithm, that isolates the ICL by modelling and subtracting the light from galaxies. \texttt{CICLE} fits galaxies using mathematical bases composed of Chebyshev rational functions and Fourier series \citep[CHEFs, ][]{jimenez-teja2012}, which are very flexible and capable of modelling galaxies with any morphology. However, objects with sharp features, such as saturated stars, diffraction spikes, or objects cut in the borders of an image, lie outside the space of elements that can be fitted by CHEFs. Although we generated PSF-subtracted images (see Sect. \ref{sec:psf}), they still contain diffraction spikes that CHEFs cannot fit (Fig. \ref{fig:rgb_a2390}). For this reason, we first mask all the stars located within the A\,2390 field, then later run \textsc{SExtractor} to detect the galaxies and, finally, model and remove them with \texttt{CICLE}.\\

The modelling of the BCG represents a challenge for any galaxy method, given its spatial coincidence, in projection, with the ICL. Indeed, the BCG is usually surrounded by a diffuse and extended halo that is complex to disentangle from the ICL. To outline the boundaries of the BCG-dominated region, \texttt{CICLE} calculates a curvature map, which expresses for each point of a surface, the local change in the slope of the surface at that point with respect to the surroundings. When the surface is the projected distribution of the composite BCG+ICL system, we can separate the two components because they usually have different slopes. \texttt{CICLE} naturally traces the transition from the BCG- to the ICL-projected distributions (or more specifically, where each one of the two distributions dominates) by identifying the curve of points where the curvature changes most.

\texttt{CICLE} operates in two dimensions, so no intrinsic assumptions on the shape or symmetry of the sources are made (either for the galaxies or the ICL). \texttt{CICLE} has been tested against simulations \citep{jimenez-teja2016}, which find that it has an error of less than 1\% in ICL measurements made on clusters between $0.2<z<0.3$, the redshift range that includes A\,2390. For A\,2390, we run \texttt{CICLE} on the cirrus- and background-removed images derived in Sect. \ref{subsubsec:multiscale_modeling} with a standard configuration. This gives each band an image containing only ICL and noise, in which the measurements in the following sections are made. A BCG+ICL map is also produced by re-inserting the BCG model into the image (see Fig.~\ref{fig:cicle_j}) and a total cluster map is created by re-inserting the models of all galaxy members. 

\subsection{DAWIS}\label{sec:dawis}

\begin{figure}
\begin{center}
\includegraphics[width=\columnwidth]{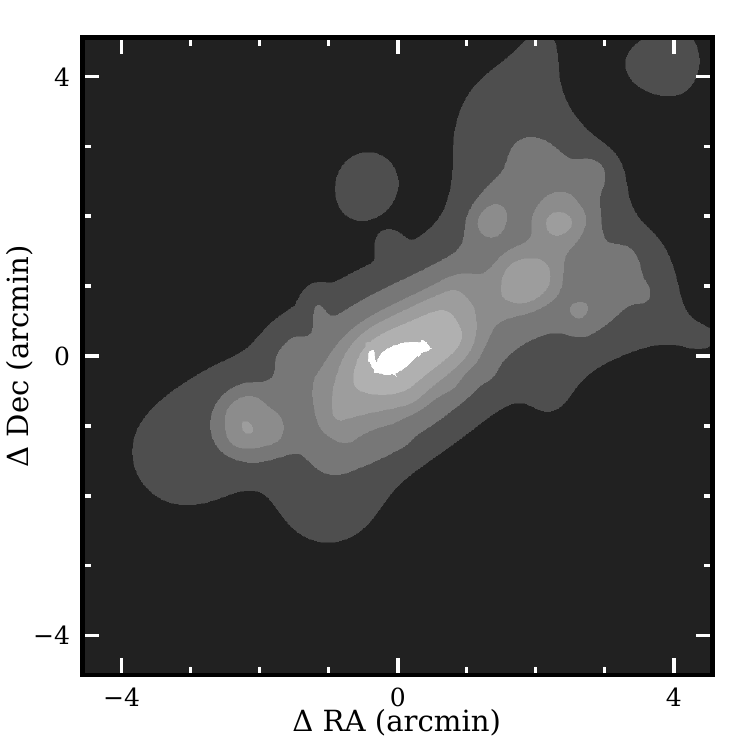}
\end{center}
\caption{A $\ang{;9.1;}\times\ang{;9.1;}$ ($2\,{\rm Mpc}\times2\,{\rm Mpc}$) region around A\,2390 on the \texttt{DAWIS} BCG+ICL \JE map.}  
\label{fig:dawis_j}
\end{figure}

A BCG+ICL map is produced using \texttt{DAWIS}, already introduced in Sect.\,\ref{subsubsec:multiscale_modeling}. The process is similar to what was described for the cirrus modelling, although adapted to ICL detection. The cirrus-corrected image of each NISP band is cropped to a box of size 
0\degf3 $\times$\,0\degf3
%$0.3\,\deg\times\,0.3\,\deg$ 
centred on the BCG, resulting in images of size $3600\,{\rm pixels}\, \times\,3600\,{\rm pixels}$. \texttt{DAWIS} is run on these images with input parameters enabling refined source modelling, with $\tau = 0.1$, $\gamma=0.5$, and a maximum number of wavelet planes of 9. Most of the ICL is modelled by selecting and summing the light profiles of atoms detected at wavelet scales greater than 5 \citep[as advocated by][]{Ellien2021} and with maximum peak coordinates inside an ellipse covering the extent of the cluster. Atoms detected at lower wavelet scales, and with maximum peak coordinates inside $5\,\arcsec$ radius around the BCG centre are also kept to model the smaller BCG core profile. The two are summed to produce for each filter the BCG+ICL model used in the rest of the analysis (the \JE band is displayed as an example in Fig.~\ref{fig:dawis_j}). Finally, the masks produced in Sect.~\ref{sec:1_D_profile} are used as priors to select atoms belonging to satellite galaxies: atoms with maximum peak coordinates within any galaxy mask are kept and their light profile is summed to produce a satellite galaxy map. A total cluster map is also produced by summing the BCG+ICL and the galaxy maps. 

\subsection{1-D profiles}
\label{sec:1_D_profile}
The third method to measure the ICL is the mask-based 1D profile fitting \citep[hereafter `Ahad', ][]{Ahad2023}, which can be used to measure the 1D BCG+ICL profile of A\,2390. The advantage of deriving the profiles in a non-parametric way is that it makes no assumptions regarding the shape of the system. Therefore, we do not assume any particular model to describe the BCG+ICL, since the results might be sensitive to the choice of the particular model and prone to degeneracies between the different parameters. 

\subsubsection{Residual background profiles}
\label{sec:bkg_profiles_1d}

Although we measure the surface brightness profiles from the PSF-cirrus-background-subtracted NISP images, there can be a low-level residual background (left-over cirrus, instrumental scattering, or zodiacal light), which can bias the ICL measurements, and is likely to have irregular distribution throughout the image. Therefore, we measure the SB profile of the residual background in 30 random locations throughout the A\,2390 FoV. The background SB profiles are measured in a similar way to that of the BCG+ICL profile (see Sect.~\ref{sec:making_1d_profiles}) centred at each random location, using the mask for all sources from Sect.~\ref{sec:mask_1d}. The random locations are generated between $2000$ and $10\,000$ pixels along both $x$- and $y$-directions to ensure that at least half of the random cutouts are always in the image. If any of the $2\,{\rm Mpc}\times2\,{\rm Mpc}$ cutouts of the 30 random background profiles have more than 50 per cent pixels masked, it is removed from the stack. Finally, the average background profile is measured. 

\begin{figure}
\begin{center}
\includegraphics[width=\columnwidth]{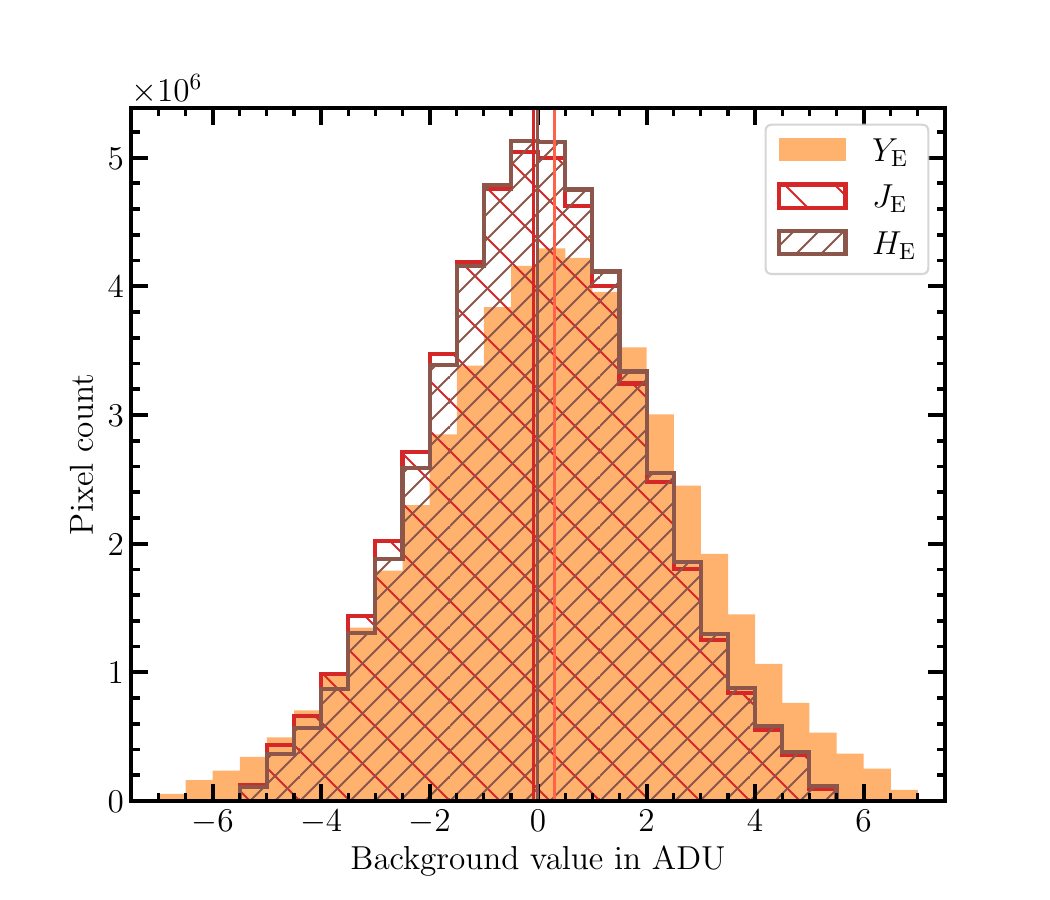}
\end{center}
\caption{Histogram of background pixels in the three cirrus-subtracted NISP bands. Because the background was also removed during the process, all three distributions are centred close to zero.}
\label{fig:bkg_hist_a2390}
\end{figure}

\begin{figure}
\begin{center}
\includegraphics[width=\columnwidth]{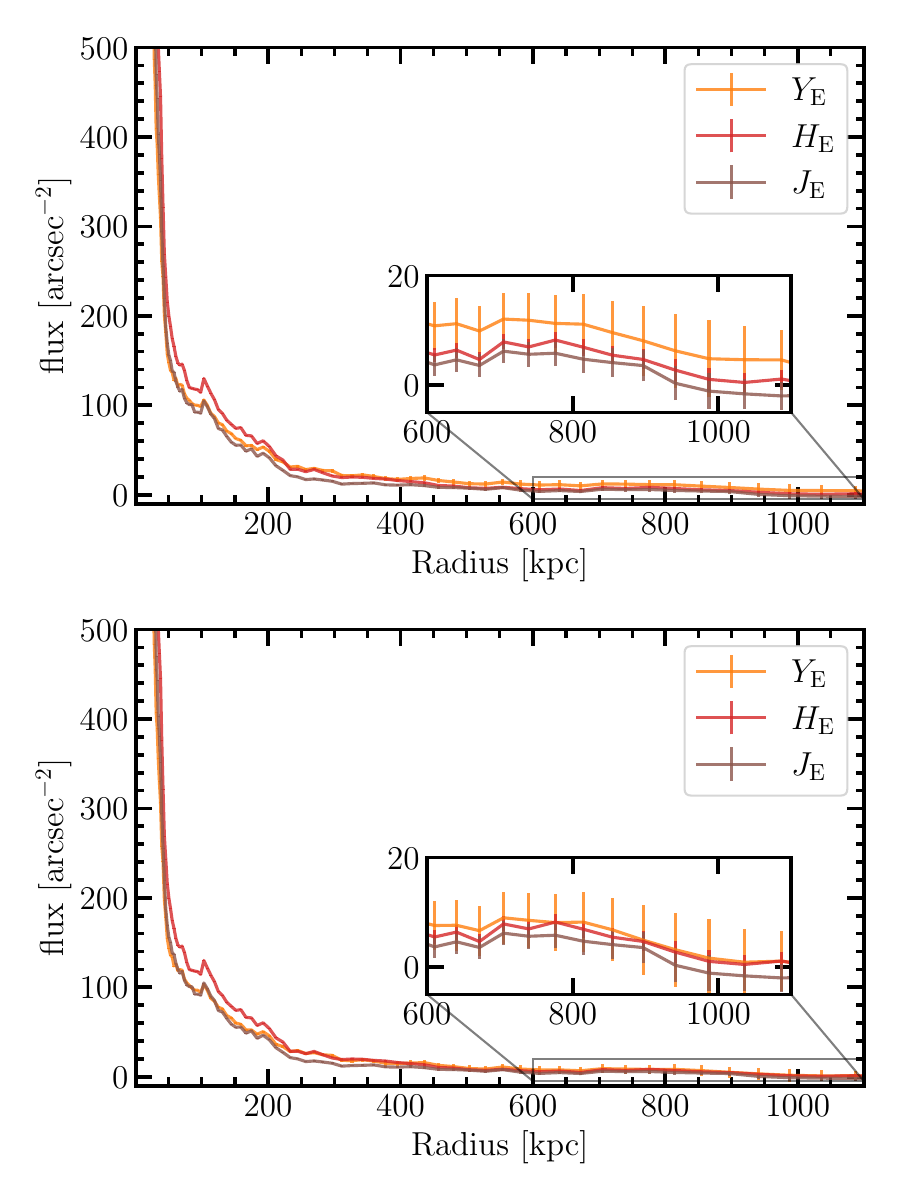}
\end{center}
\caption{Surface brightness profiles of the BCG+ICL in the three NISP bands before (top panel) and after (bottom panel) subtracting the residual background. The flux density values are shown with a magnitude zero-point of 30.}
\label{fig:flux_profiles_bkg_before_after_a2390}
\end{figure}

Figure~\ref{fig:bkg_hist_a2390} shows the distribution of the residual background pixels for the three NISP bands. As expected from the already background-removed images, their median values are within 1$\sigma$ of 0, as shown in the histogram (with median values of background pixels indicated with vertical lines in the same colours as the histograms according to the NISP filters). We note the \YE band displays a mean value slightly higher than 0, probably due to higher cirrus residuals for this filter (see Sect.~\ref{subsec:limitations_and_uncertainties}). This is also highlighted in Fig.~\ref{fig:flux_profiles_bkg_before_after_a2390}, where the maximum difference in the flux density profile of the BCG+ICL before (top panel) and after (bottom panel) the residual background removal is in the \YE band. 

\subsubsection{Measuring surface brightness profiles}
\label{sec:making_1d_profiles}

\begin{figure}
\begin{center}
\includegraphics[width=\columnwidth]{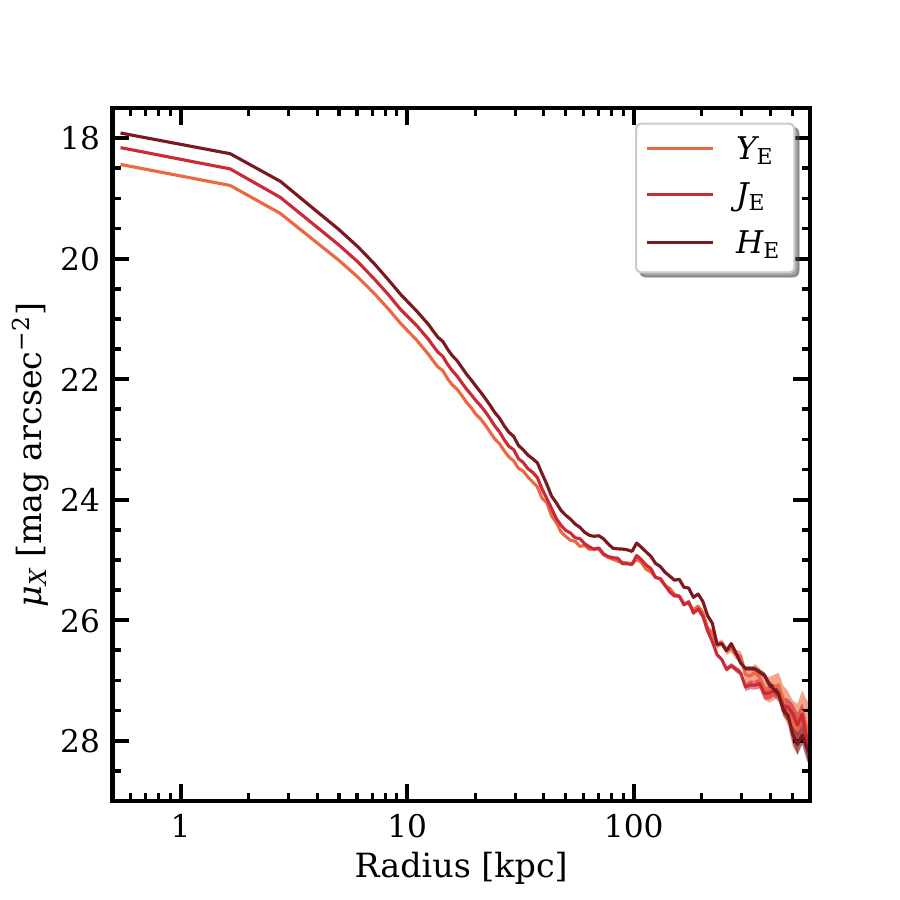}
\end{center}
\caption{Surface brightness radial profiles for the BCG+ICL of A\,2390 from the 1-D method Ahad. The plot shows the profiles for the \YE (orange), \JE (red), and \YE (dark red) NISP bands. }
\label{fig:SB_a2390}
\end{figure}

The 1-D azimuthally averaged profiles are created using circular apertures centred at the BCG of A\,2390 out to 1\,Mpc radial distance. The central parts of the profiles are linearly binned from one to 10 pixels for better sampling; logarithmic binning is used beyond that. Masks (ii) and (iii) from Sect.~\ref{sec:mask_1d} are used to calculate the BCG+ICL and total cluster light profiles from the unmasked pixels, respectively. The average residual background profiles (Sect.~\ref{sec:bkg_profiles_1d}) are subtracted from the BCG+ICL and total cluster light profiles. Finally, the surface brightness radial profiles, shown in Fig.~\ref{fig:SB_a2390}, are converted to $\mathrm{mag\,arcsec}^{-2}$. 

\section{\label{sc:Results} Results}

\subsection{Surface brightness radial profiles of BCG+ICL}

We derive the radial profiles in each of the three NISP bands following a similar methodology as in \citet{Ahad2023}, as described in Sect.~\ref{sec:making_1d_profiles}. The 1-D radial surface brightness profiles as a function of radius for the \YE\ (orange), \JE\ (red), and \HE\ (dark red) bands are shown in Fig. \ref{fig:SB_a2390}.  

We conservatively decide not to explore the ICL beyond 600~kpc, since the uncertainties become too large ($>0.5$ mag arcsec$^{-2}$). The BCG+ICL profile roughly follows a \citet{Sersic1968} profile, out to 50\,kpc where we start seeing more structure in the profiles. This is probably due to some unmasked light from satellite galaxies, which is already too dispersed and mixed into the ICL for us to mask it properly. 
Around 70\,--\,80\,kpc, we see an excess in the profile's extension in the inner 50 kpc, a signature of accreted material, i.e. the ICL.

\subsection{ICL fractions}\label{sec:fractions}

Measuring the amount of light in the ICL with respect to the total light of the cluster (brightest cluster galaxy, satellite galaxies, and ICL), the ICL fraction, enables us to evaluate the efficiency of the processes that shaped the cluster. Exploring the BCG+ICL fraction (the amount of light in the BCG+ICL with respect to the total light of the cluster) can give us clues about their common evolution.

Table~\ref{table:fractions} shows the BCG+ICL and ICL fractions (and luminosities) of A\,2390 derived with the three methods presented in this work, namely, \texttt{CICLE}, \texttt{DAWIS}, and the 1-D profile Ahad method. The ICL and BCG+ICL fractions are derived independently in each method, using their respective ICL, BCG+ICL, and total cluster light integrated flux values. For consistency, all integrated flux values are computed within 600\,kpc from the BCG centre 

All methods use the same way to separate the BCG from the ICL: cutting out the inner 50\,kpc of the BCG, as in Bellhouse et al. (in prep.), to produce results compatible with future \Euclid\ ICL studies. For the Ahad method, the BCG+ICL and total cluster (BCG+ICL+satellites) fluxes are computed by integrating their respective profiles, out to 600\,kpc. To measure the ICL flux, we integrated the BCG+ICL 1-D profile from 50 to 600\,kpc. For the \texttt{CICLE} and \texttt{DAWIS} methods, the BCG+ICL fluxes are computed by integrating the respective BCG+ICL maps, out to 600\,kpc. The ICL flux is computed by integrating the BCG+ICL maps from 50 to 600 kpc. In the same way, the total cluster flux is derived from their respective total cluster maps (as described in Sect.~\ref{sec:cicle} for \texttt{CICLE} and Sect.~\ref{sec:dawis} for \texttt{DAWIS}) integrating out to 600\,kpc.

Although a detailed comparison between the methods is beyond the scope of this paper \cite[for more information on this see][]{Brough2024}, we see that the fractions derived are roughly similar. The average BCG+ICL fraction is 29\,$\%$, for all bands and all methods, and the average ICL fraction is 24\,$\%$. Note that the \texttt{CICLE} fractions appear slightly higher than \texttt{DAWIS} and Ahad. However, this difference is only a few per cent, and compatible within uncertainties.  

The BCG+ICL fractions quoted here (29\,$\%$ for all bands and methods) are in agreement with those in the literature for clusters at the redshift of A\,2390 \citep[$z\sim 0.23$, ][]{zhang19, Furnell2021, Sampaio-Santos2021}. It also agrees with the fractions from \citet{gonzalez05}, \citet{kluge21}, and \citet{Ragusa2023} of nearby massive clusters.
The ICL fraction is on average $24$\,$\%$, which agrees with the estimates of \citet{Burke2015}, \citet{Furnell2021}, \citet{zhang2024}, and \citet{golden-marx2024} although it is towards the higher end. This could be because A\,2390 is a very massive cluster ($M_{200,{\rm c}} = 1.6\times10^{15}$ \si{\solarmass}). The small difference between the ICL and BCG+ICL fraction ($24$\,$\%$ versus $29$\,$\%$), suggests that most of the light in the BCG+ICL in this cluster is in the ICL. The fraction of the ICL over the BCG+ICL component is on average $\sim$\,80\,$\%$, which is high compared to the values for other clusters (typically around 65--70\,\%, see figure 5 in \citealt{Montes2022}). This higher fraction is remarkable considering we do not include the ICL light in projection to the BCG in the inner 50\,kpc, making it a lower limit.

\begin{table*}[t]
\centering
\caption{ICL and BCG+ICL fractions and luminosities of A\,2390, measured by the different methods explored in this work, namely \texttt{CICLE}, \texttt{DAWIS}, and 1-D Profile Ahad.}
\tabcolsep=0.4cm
\begin{tabular}{ccccccc}
\hline
\\[-9pt]
 & \multicolumn{2}{c}{CICLE} & \multicolumn{2}{c}{DAWIS} & \multicolumn{2}{c}{1-D Profile Ahad}\\
\hline
& & & & & \\[-9pt]
   & $f_{\rm{ICL}}$ & $f_{\rm{BCG+ICL}}$ & $f_{\rm{ICL}}$ & $f_{\rm{BCG+ICL}}$ & $f_{\rm{ICL}}$ & $f_{\rm{BCG+ICL}}$  \\
& & & & & \\[-9pt]
$\YE$ & $0.36\pm0.05$ & $0.41\pm0.05$  & $0.25\pm0.03$ & $0.32\pm0.02$  & $0.25\pm0.01$ & $0.30\pm0.01$  \\
& & & & & \\[-9pt]
$\JE$& $0.24\pm0.03$ & $0.29\pm0.04$ & $0.23\pm0.02$ & $0.30\pm0.02$ & $0.18\pm0.01$ & $0.23\pm0.01$ \\ 
& & & & & \\[-9pt]
$\HE$ & $0.24\pm0.03$ & $0.28\pm0.03$ & $0.21\pm0.01$ & $0.30\pm0.02$ & $0.18\pm0.01$ & $0.23\pm0.01$  \\ 
\hline %%%%%%%%%%%%
& & & & & \\[-9pt]
  & $L_{\rm ICL}$ & $L_{\rm BCG+ICL}$ & $L_{\rm ICL}$  & $L_{\rm BCG+ICL}$ & $L_{\rm ICL}$ & $L_{\rm BCG+ICL}$ \\
& & & & & \\[-9pt]
  & $(10^{12}L_{\odot})$ & $(10^{12}L_{\odot})$ & $(10^{12}L_{\odot})$  & $(10^{12}L_{\odot})$ & $(10^{12}L_{\odot})$ & $(10^{12}L_{\odot})$ \\
& & & & & \\[-9pt]
$\YE$ & $2.26\pm0.16$ & $2.51\pm0.18$ & $1.07\pm0.08$ & $1.52\pm0.09$ & $1.92\pm0.03$ & $2.46\pm0.03$ \\
& & & & & \\[-9pt]
$\JE$ & $1.57\pm0.15$ & $1.86\pm0.16$ & $1.52\pm0.10$ & $2.06\pm0.12$ & $1.65\pm0.01$ & $2.30\pm0.01$ \\ 
& & & & & \\[-9pt]
$\HE$ & $1.96\pm0.18$ & $2.25\pm0.19$ & $1.44\pm0.11$ & $2.32\pm0.12$ & $2.05\pm0.01$ & $2.88\pm0.01$  \\ 
%L_ICL Ahad: 14.03 ,  14.26, 14.04
\hline
\end{tabular}
\label{table:fractions}
\end{table*}

\subsection{Colours of the ICL and satellite galaxies}

Radial colour gradients constrain the physical processes that build up the ICL, and consequently, the BCG \citep[][]{Zibetti2005, MT14, DeMaio2015, Spavone2020, Montes2021, Ragusa2021, Ragusa2022, golden-marx2023}. Each process, such as tidal stripping or galaxy mergers, leaves a distinct imprint on the stellar population reflected in its spectra and the measured colours. For this reason, comparing the colour of the ICL with that of the satellites reveals clues about its progenitor population.

\begin{figure}
\begin{center}
\includegraphics[width=\columnwidth]{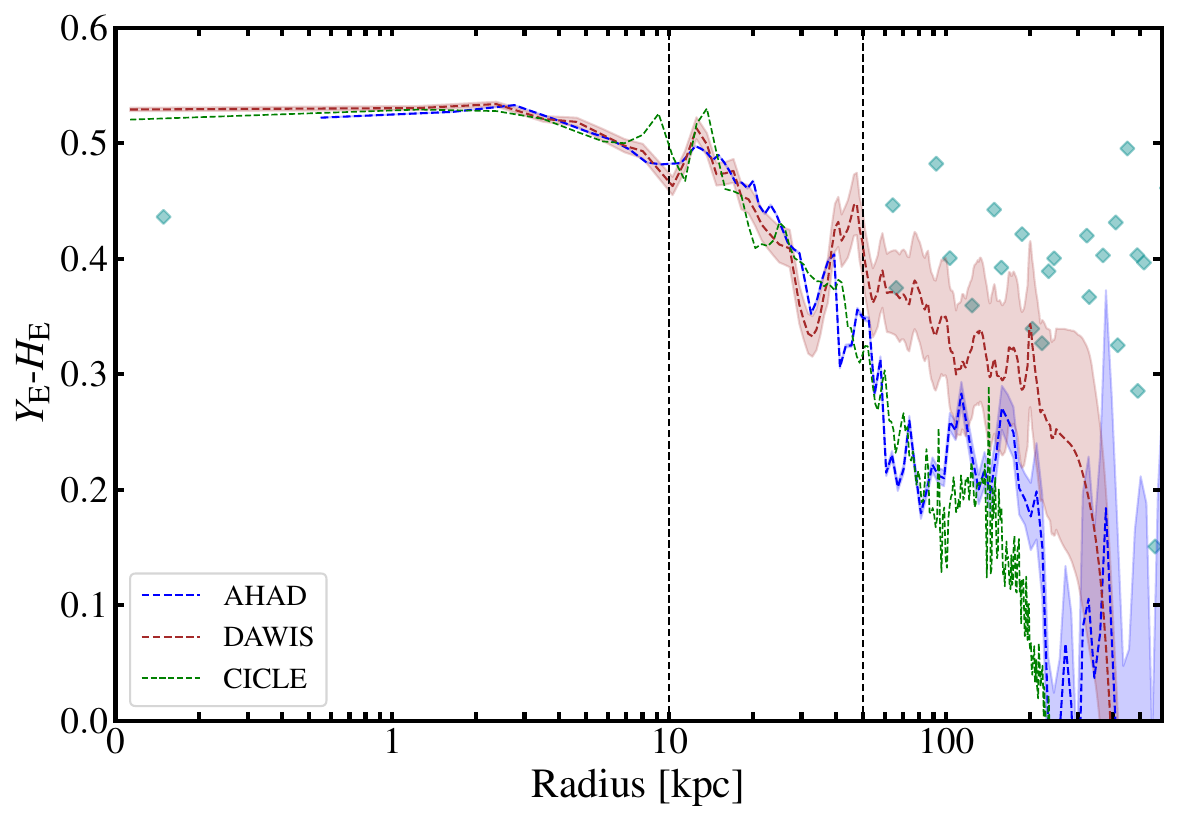}\\
\end{center}
\caption{Colour radial profiles for the BCG+ICL system of 
A\,2390. Green diamonds represent the cluster selected members, while red, blue, and green lines are the ICL radial profiles estimated with the three methods presented in Sect. \ref{sec:Method}: Ahad in blue; \texttt{CICLE} in green; and \texttt{DAWIS} in red. The vertical lines indicate 10 and 50 kpc, to guide the eye. }
\label{fig:radial_color}
\end{figure}

Figure \ref{fig:radial_color} shows the radial $\YE-\HE$ colour profile of the BCG+ICL of A\,2390. We use this colour to study the stellar populations of the BCG+ICL system because it covers a wider wavelength range than any of the other colours. In Fig.~\ref{fig:vazdekis} in Appendix \ref{app:models}, we show the \Euclid NIR colour from the \citet{Vazdekis2016} stellar population models. %The $\YE-\HE$ colour covers a wider range of values than other colours, enabling better identification of BCG+ICL stellar population changes.

We derive colour profiles for the three methods described in Sect.~\ref{sec:Method}. For the 1-D profile method, we subtract the profiles derived in Sect.~\ref{sec:1_D_profile}.
The radial surface brightness profiles, and therefore colour profiles, for \texttt{DAWIS} and \texttt{CICLE} are derived in a similar way as in the Ahad 1-D profile method: using circular annular apertures on the 2-D BCG+ICL maps that result from the two methods described in Sect.~\ref{sec:Method}. We use small steps in radial distance to better trace the ICL's behaviour.

In Fig.~\ref{fig:radial_color}, we overlay the colours for the cluster members (light green diamonds), which are computed using the photometric values from the member catalogue presented in Sect.~\ref{sec:members}.  
The MW extinction corrections for each \Euclid filter are derived using the \textit{Planck} thermal dust map \citep{Planck_dust2014,Gordon2023} extinction law, and assuming a spectral energy distribution of a $5700$\,K blackbody (following the same method as described in \citealp{Kluge2024}).

The cluster members do not show any radial dependence in their colour from the core to the outskirts (though we note that this is probably linked to the cluster member selection method). The BCG and cluster members have a mean colour of $\YE-\HE =0.39$. Cluster member colour histograms are shown in Appendix \ref{app:members}.  
The colour profiles of the BCG+ICL present two distinct regions: flat in the core out to 5 kpc; and a negative slope from 5 to around 450\,kpc. 
The flat colour profile at $<$\,5\,kpc is consistent with a mixing of the stellar populations in the centre of the BCG, while the negative colour gradient at $>$\,5\,kpc indicates a gradient in the stellar populations of the BCG+ICL system: they become bluer with radius. Because the $\YE-\HE$ colour mainly traces a change in metallicity (Appendix \ref{app:models}), this indicates that the stellar populations become metal-poor with radius due to the accretion of smaller systems into the BCG. We also note that because of modelling uncertainties, the \texttt{DAWIS} colour profile is redder at large radii ($>$\,200\,kpc) than seen in the other two methods.

\subsection{The spatial distributions of A\,2390 cluster members}

As discussed, colour gradients indicate that the ICL is at least partially formed through tidal stripping of satellite galaxies. While some of this occurs when satellites fall into the cluster, tidal stripping also occurs when satellite galaxies interact with the BCG. Therefore, comparing the spatial distribution of the ICL to that of the satellites may inform our understanding of the population of galaxies that form the ICL.  

\subsubsection{Galaxy density map}
\label{subsec:galaxy_density_map}
\begin{figure}
\begin{centering}
\includegraphics[width=0.49\textwidth]{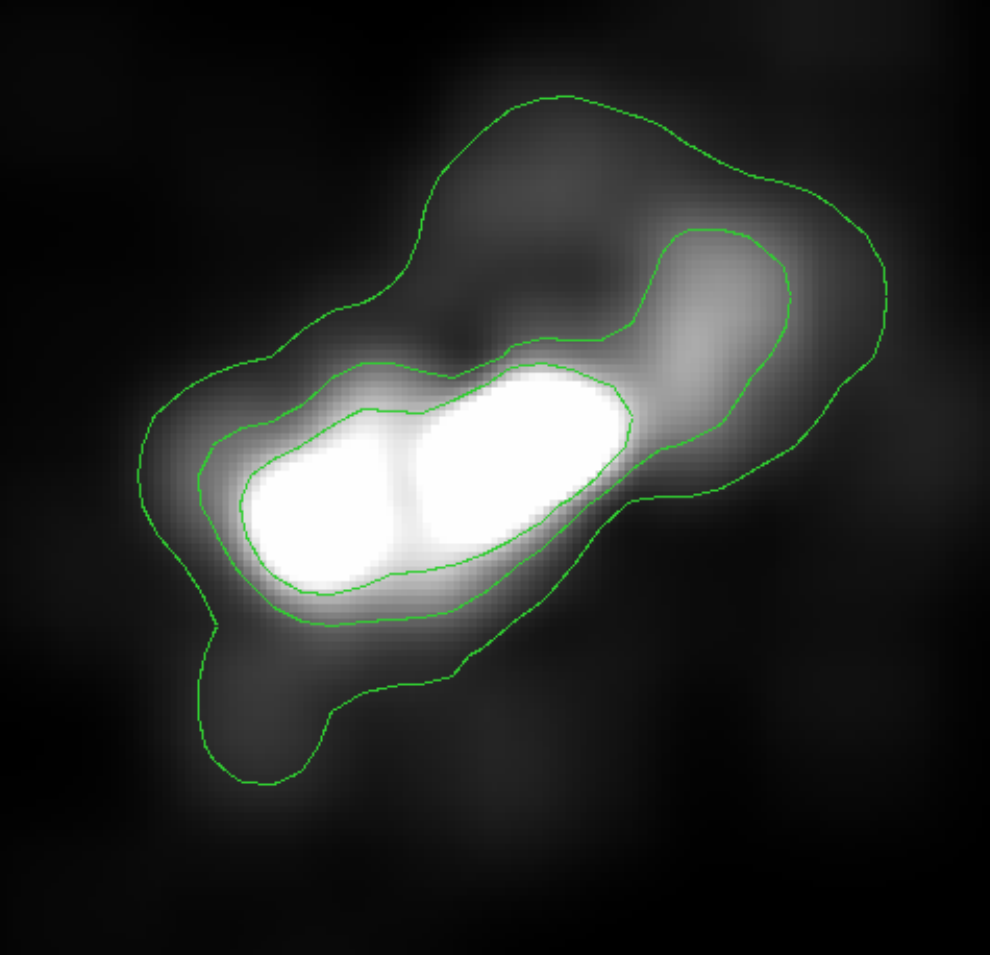}
\end{centering}
\caption{Density map of the 219 galaxies selected as belonging to the cluster A\,2390 (see main text). The green contour levels correspond to 3$\sigma$, 10$\sigma$, and 20$\sigma$ above the background.  North is up and east is to the left. The image size is  $\ang{0.4;;}\,\times\,\ang{0.4;;}$ (5.24\,Mpc\,$\times$\,5.24\,Mpc) and is centred on the BCG. }
\label{fig:densmap}
\end{figure}
To investigate this, we first look at the spatial distribution of the entire sample of 219 potential cluster members described in Sect.~\ref{sec:members} by measuring a density map, as shown in Fig.~\ref{fig:densmap}. For this density map, we use the adaptive kernel technique with a generalised Epanechnikov kernel \citep{Silverman86}. This implementation is summarised in \cite{Dantas+97}, based on an earlier version developed by Timothy Beers (ADAPT2) and improved by
\cite{Biviano+96}.  The statistical significance is established by bootstrap resampling of the data.  We perform 100 bootstrap realisations with a pixel size of \ang{;;5.4} and the density map is computed for each realisation. For each pixel of the final bootstrap map, the value is taken as the mean over all realisations \citep[see][]{Durret+16}. 

We derive the significance level of our detection by estimating the mean value and dispersion of the background. For this, we draw histograms of the pixel intensities and fit them with a Gaussian, as illustrated in \cite{Durret+16}.  The mean value of the Gaussian gives the mean background level, and the width of the Gaussian is the dispersion, $\sigma$. We then compute the contour values corresponding to $n\sigma$ detections as the background plus $n\sigma$.

In Fig.~\ref{fig:densmap}, we see an elongated distribution of the satellites with a similar shape and ellipticity to what is seen for the ICL in Fig.~\ref{fig:dawis_j}.  Interestingly, we appear to see two distinct populations, the main cluster and a smaller secondary population in the south-east. This distribution does not change if we only include the spectroscopic members.

\subsubsection{Galaxy morphology distribution}
\label{subsec:morph}
While Fig.~\ref{fig:densmap} suggests a similar spatial distribution between the ICL and the satellites, it does not tell us whether all satellite galaxies trace the ICL. Each cluster member's morphology (using the sample described in Sect.~\ref{sec:members}) is visually classified by two of the authors (JBGM, PD) as either elliptical or spiral using the \Euclid VIS images. Although some member galaxies have previous morphological classifications \citep{Abraham1996}, \Euclid's higher resolution improved the accuracy of these classifications. 

\begin{figure}
\begin{center}
\includegraphics[width=0.49\textwidth]{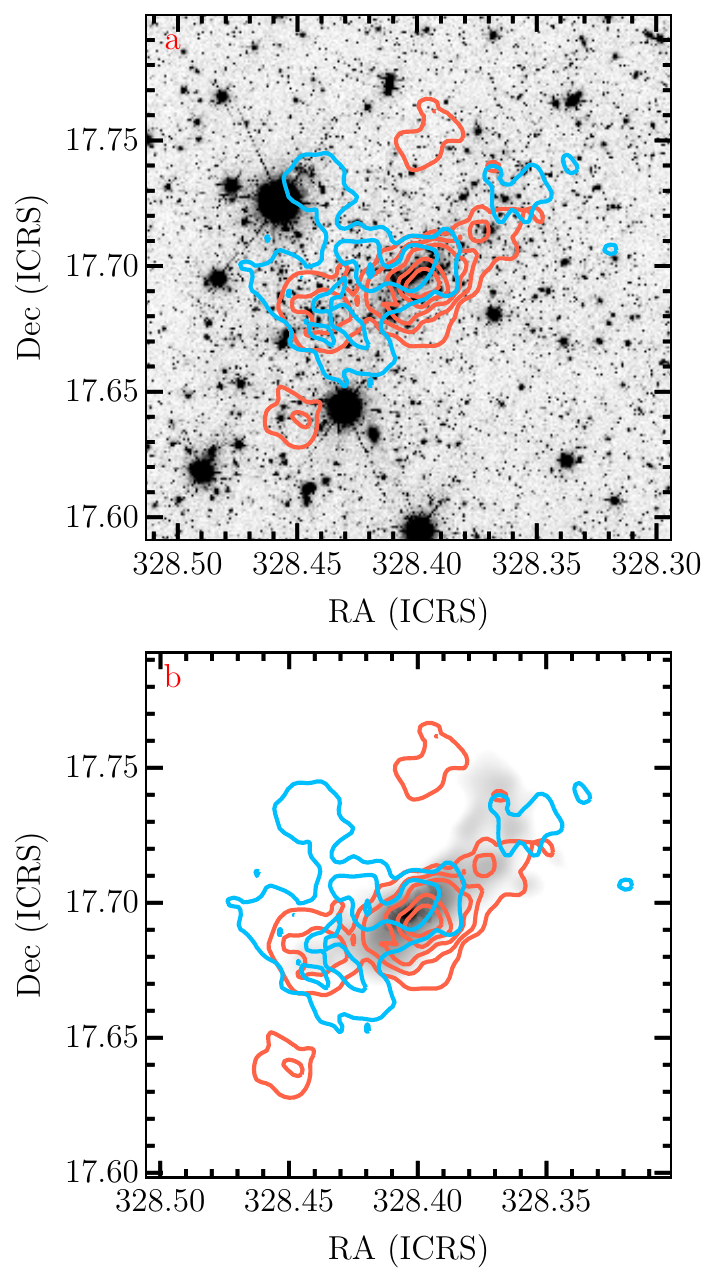}
\end{center}
\caption{Contour maps of the spatial distribution of elliptical (red) and spiral (blue) galaxies overlaid on the \HE-band image of A\,2390 in the upper panel, and the \texttt{DAWIS} BCG+ICL map in the lower panel. The figure shows a region of $\ang{;;700}\times\ang{;;700}$ ($2.5~{\rm Mpc}\times2.5~{\rm Mpc}$) centred on the BCG.}
\label{fig:morph_contours}
\end{figure}

For the analysis shown in Fig.~\ref{fig:morph_contours}, we focus on the 129 members of A\,2390 within a projected distance of 1\,Mpc of the BCG \citep[corresponding roughly to half the virial radius; ][]{Li2009} and divide our sample into spirals (shown in blue) and ellipticals (shown in red). Although these cluster members all lie upon a tight red sequence (see Fig.~\ref{fig:CMD}), not all are morphologically elliptical; a number have spiral features only visible due to \Euclid's high spatial resolution. For each population, we oversample the data and measure the number of galaxies within \ang{;;48.96} (ten times the spatial resolution of the NISP data used to measure the ICL) of every point separated by \ang{;;19.58} (4 times the spatial resolution of NISP).  We then construct contours to identify the spatial distribution of the spiral and elliptical populations.

In Fig.~\ref{fig:morph_contours}, we show the contours of the spirals (in blue) and ellipticals (in red) onto the \HE-band image of A\,2390 in the upper panel and the \texttt{DAWIS} BCG+ICL map in the lower image.  These contours allow us to directly compare how the distribution of morphological populations of satellite galaxies align with the cluster (upper panel) and BCG+ICL (lower panel). In agreement with \citet{Abraham1996} and our understanding of red sequence formation \citep[e.g.,][]{gladders00, rykoff2014}, these contours show that the central region of the cluster is predominantly populated by elliptical (red) galaxies.
Moreover, the comparison between the satellite galaxy distributions and the \texttt{DAWIS} BCG+ICL model in Fig.~\ref{fig:morph_contours} illustrates that the elliptical galaxies (red) cause the alignment between the ICL and satellite population.
The spiral spatial distribution (blue) on the other hand is more scattered and is shifted toward the south-west -- it does not trace the entire ICL but rather substructures.
We note that the satellite galaxy and ICL distributions identify similar features, most noticeably the asymmetry in the north-west direction.
The elliptical satellite contours, similar to Fig.~\ref{fig:densmap}, identify a secondary population of galaxies in the south-east direction. This subgroup appears to have more spiral galaxies than the core.  However, the most significant observation is that the contours of the elliptical galaxies suggest that the galaxy distribution centroid is offset from the BCG by about 20$\arcsec$ ($\approx$\,70\,kpc).

\subsubsection{Detection of substructures} \label{sec:substructures}

We apply the DS+ algorithm \citep{benavides2023} to detect groups among the 184 cluster members spectroscopically identified in Sect.~\ref{sec:members}. DS+ is an updated version of the \cite{dressler1988} algorithm, which compares the local velocity field around each galaxy with that of the whole cluster. Statistically significant departures of the two distributions are indicative of the presence of a group. DS+ also identifies the galaxies that compose the groups with a statistical approach. Under the `non-overlapping' mode, this identification is unique, meaning each galaxy is assigned to a single group and cannot belong to several groups simultaneously. 

\begin{figure*}
\begin{center}
\includegraphics[width=0.49\textwidth]{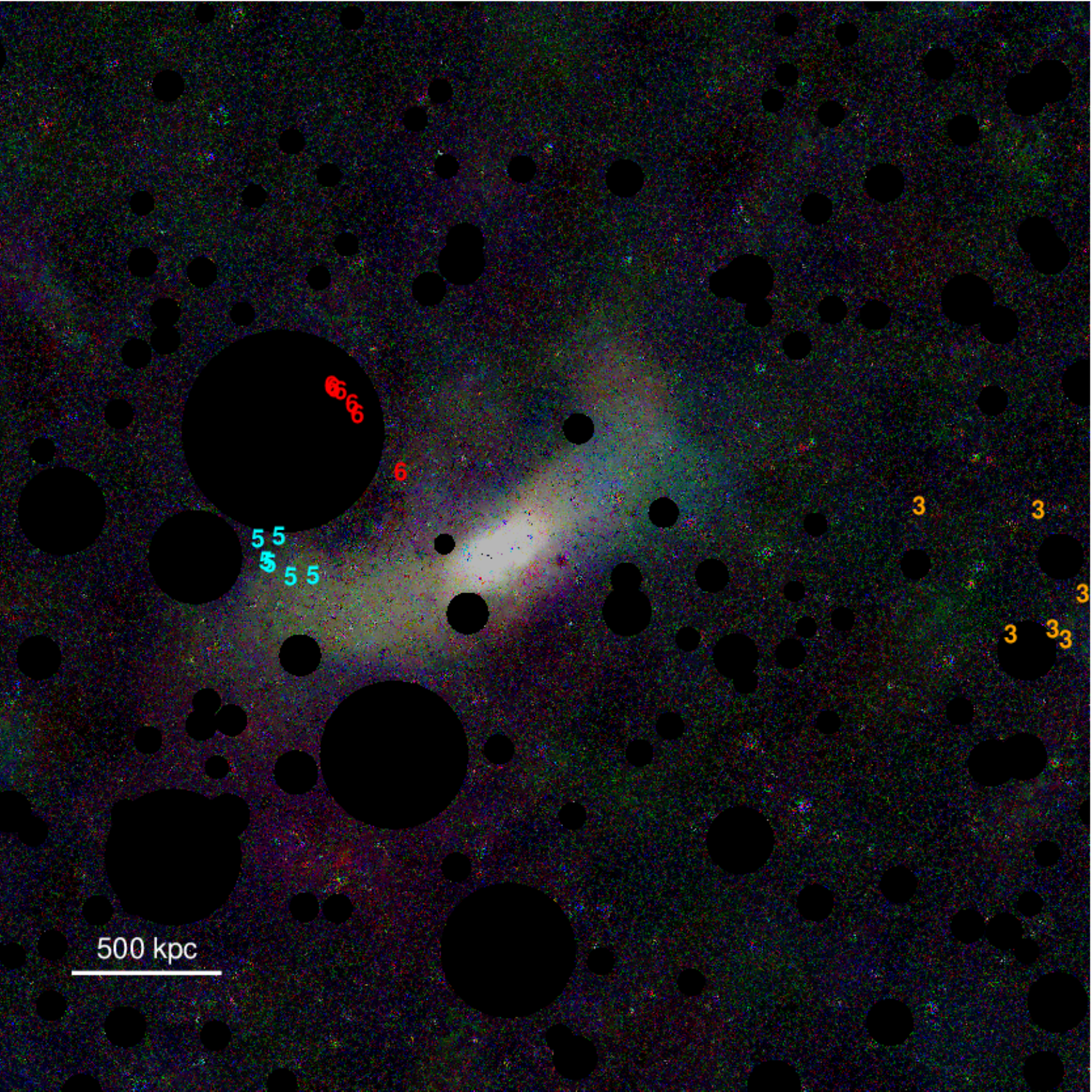}
\includegraphics[width=0.49\textwidth]{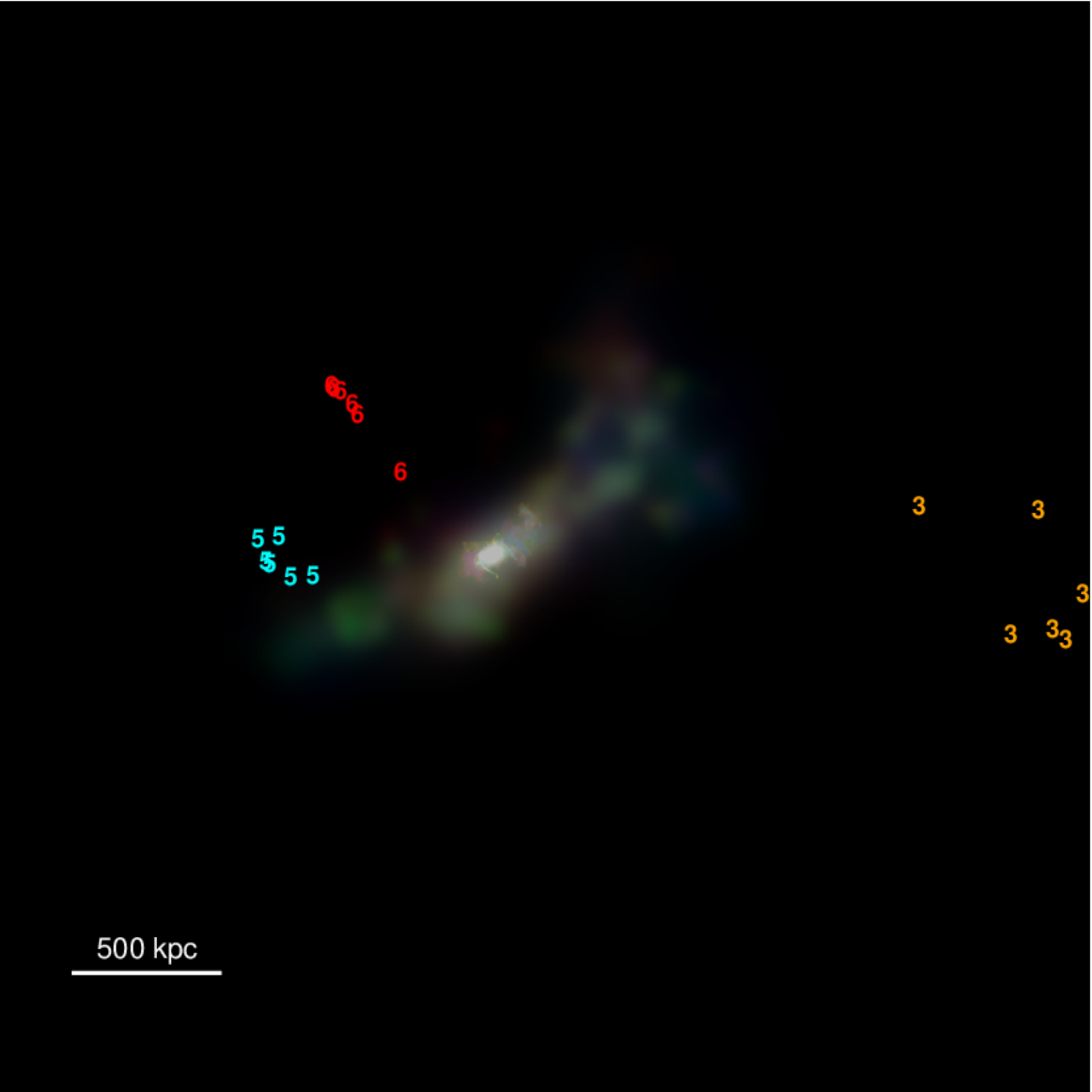}
\end{center}
\caption{RGB images of the BCG+ICL in the core of A\,2390 ($700\,\arcsec\times 700\,\arcsec$). The left panel shows the \texttt{CICLE} RGB map and the right panel the \texttt{DAWIS} RGB map. The numbers in both maps mark the position of the galaxies in the groups identified by DS+ (see Sect. \ref{sec:substructures}), listed in Table~\ref{table:DS+}. Both colour images were generated with Trilogy \citep{coe2012}.}
\label{fig:rgb_groups}
\end{figure*}

We set DS+ with a probability threshold of 0.1 and ran it with 1000 simulations in the non-overlapping mode, as recommended in \cite{benavides2023}. We identified 6 groups with more than four members; their positions are shown in Fig.~\ref{fig:rgb_groups} and their properties are described in Table~\ref{table:DS+}. Note that groups 1, 2 and 4 fall outside the field of view shown in Fig.~\ref{fig:rgb_groups}.

Figure~\ref{fig:rgb_groups} shows RGB maps from \texttt{CICLE} (left) and \texttt{DAWIS} (right) created from the BCG+ICL maps for each of the NISP bands. As shown in Fig.~\ref{fig:rgb_groups}, the different groups follow the overall ICL distribution. In particular, groups 5 and 6 are associated with the south-eastern ICL structure, whereas group 3 may be infalling from the west.\footnote{We assume that the diffuse light and the galaxies are linked by how the ICL forms, but it may be the case that some of the galaxies are along the line of sight and not at the distance of the diffuse light.} %Interestingly, there is a patch of ICL that does not seem associated with any group (between groups 1 and 2). However, in Fig.~\ref{fig:morph_contours} there are some galaxies in this region, suggesting that the progenitor of this ICL patch may be a small group ($<$\,4) of galaxies.\footnote{Groups of less than four galaxies have low statistical significance in detection algorithms.}

The colours of the ICL patches in Fig.~\ref{fig:rgb_groups} are not homogeneous, a sign that the ICL in these groups is not well-mixed yet. 
%For example, the `orphan' patch described above is redder than the others in \texttt{CICLE} and \texttt{DAWIS} RGB maps. 
%The ICL associated with group 3, on the other hand, appears bluer, while groups 5 show an intermediate colour (green). 
For example, the ICL in the north-western direction appears bluer, while the ICL associated with group 5 shows an intermediate colour (green).
Nevertheless, these patches have a similar characteristic size as some of the cirrus filament residuals seen in other areas of the cirrus-corrected images (see Sect.~\ref{subsec:limitations_and_uncertainties} and Fig.~\ref{fig:cirrus_residuals}). We do not rule out that these colours might be affected by cirrus contamination.

\begin{table}
\centering
\caption{Properties of the 6 groups identified by the DS+ method: group identifier, number of members, distance to the centre of the cluster, velocity dispersion, and mean velocity.}
\begin{tabular}{ccrrr}
 \hline
\\[-9pt]
ID & $N_{\rm gal}$ & Dist\,(kpc)&  $\sigma$\,(\SI{}{\km\per\second})&  $v_{\rm{mean}}$  \,(\SI{}{\km\per\second})\\
\hline
\\[-9pt]
1 & 12 & 3094 & 1557 & 67\,255 \\
2 &  6 & 2208 &  288 & 69\,725 \\
3 &  9 & 1923 &  785 & 67\,384 \\
4 &  9 & 4073 &  517 & 69\,461 \\
5 &  6 &  711 &  898 & 70\,074 \\
6 &  6 &  665 & 1896 & 67\,232 \\
\hline
\end{tabular} 
\label{table:DS+}
%1 & 6 & 1051.73 & 17\,469.73 & 101\,470.35 \\
%2 & 9 & 628.34 & 19\,478.69 & 86\,123.31\\
%3 & 6 & 586.85 & 26\,411.37 & 89\,796.78 \\
%4 & 6 & 493.48 & 11\,356.74 & 84\,685.37 \\
%5 & 6 & 607.94 & 12\,216.76 & 88\,528.26 \\
%6 & 6 & 603.71 & 13\,000.44 & 93\,875.36 \\
%7 & 6 & 1569.70 & 813.49 & 67\,554.23 \\
%8 & 9 & 2304.57 & 1136.55 & 68\,240.42 \\
%9 & 6 & 1236.19 & 26\,958.18 & 92\,404.02 \\
%10 & 6 & 2603.34 & 646.60 & 66\,980.12 \\
%11 & 9 & 4073.31 & 517.20 & 69\,461.21 \\
\end{table}

\subsection{Comparison with total mass and X-ray maps}\label{sec:mass_xrays}

One of the primary goals of \Euclid is to map the mass distribution in the Universe using weak gravitational lensing measurements. The ICL is an effective tracer of dark matter in clusters \citep{MT19, Alonso-Asensio2020, Sampaio-Santos2021, Yoo2024}. Therefore, the ICL provides an additional constraint to improve these derived mass distributions in clusters.
In addition, the comparison of ICL, X-ray, and mass map distributions can give us clues about the dynamical state of the cluster \citep[e.g.,][]{Kluge2024}. 

Here, we compare the two-dimensional distribution of dark matter in A\,2390 (as traced by gravitational lensing models) with the X-ray and BCG+ICL (DAWIS \HE) maps. For this, we follow the steps outlined in \citet{MT19}, summarised below.

We use the mass map of A\,2390 as derived in Diego et al. (2024, in prep.). The map is a joint strong+weak lensing solution using WSLAP+ \citep{Diego2005, Sendra2014}. WSLAP+ is a free-form code, that is, it does not assume any mass distribution. It improves the \citet{EROLensdata} results by adding new photometric redshift estimates for the galaxies in the \Euclid FoV. To compare the shape of the cluster X-ray emission to the mass and ICL distributions, we also retrieve Advanced CCD Imaging Spectrometer (ACIS) images of A\,2390 from the Chandra Data Archive.\footnote{\url{http://cda.harvard.edu/chaser/}} The X-ray maps are from Observation 4193 (PI: S. Allen). 

We derive the isocontours for the three components: BCG+ICL; total mass; and X-rays. For a sensible comparison, the isocontours of each map are obtained at the same physical radial distances: 50, 100, 150, 200, 300, 500, 600, and 700\,kpc. To do that, we derive radial profiles of the three components. Note that the purpose of these radial profiles is to obtain an intensity at a given radial location to derive the isocontours for each of the three components.  

\begin{figure}
\begin{center}
\includegraphics[width=0.51\textwidth]{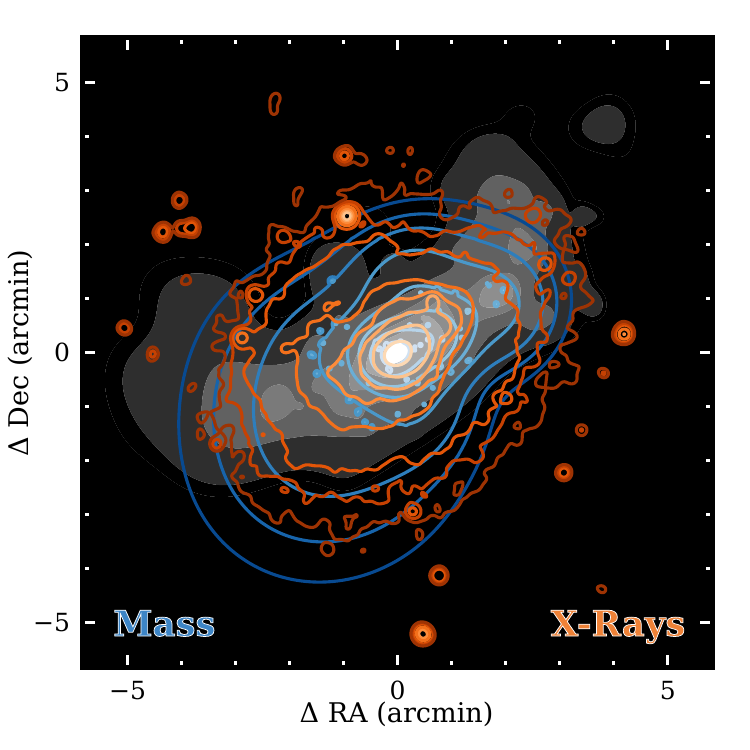}
\end{center}
\caption{Region of $700\arcsec\times700\arcsec$ around A\,2390. The background is the ICL filled-contour map from \texttt{DAWIS} (\HE\ band) while the blue and orange lines are the mass and X-ray contours, respectively. The contours are at 50 (lightest shade), 100, 150, 200, 300, 500, 600, 700\, kpc (darkest shade).}
\label{fig:mass_xrays}
\end{figure}

We assume the centre for each of the maps to be the location of the BCG, since the peaks of the mass map and the X-rays are also located there.
Then, we obtain the radial profiles of the BCG+ICL, X-ray emission, and mass. The profiles are constructed averaging over circular bins out to 1\,Mpc. Once the intensities at the different radial distances are obtained, we use \texttt{contour} in \texttt{matplotlib} to obtain the isocontour lines. Figure~\ref{fig:mass_xrays} shows the comparison between the contours (shapes) of the different components. The black and white background is the \texttt{DAWIS} BCG+ICL map, while the contour lines show the X-rays (orange) and the mass (blue) distributions. 

At the very centre of the cluster, within 300\,kpc, the three components are very similar to each other. Beyond 300\,kpc, however, the three components begin to diverge. The X-ray contours show an asymmetry towards the south-east at about 300\,kpc, although they become more regular again at larger distances. 
The BCG+ICL map appears more elongated (elliptical) than the mass and X-ray maps. The mass map is elongated along the north-east to south-east axis, as in the BCG+ICL map. Interestingly, towards the south-east the mass contours widen perpendicular to the axis, probably indicating the presence of the south-east group of galaxies. The same happens, although to a lesser extent, to the north-east. 
Al larger radii, the contours are rounder than the diffuse light, more similar to the X-rays. Therefore, it appears that the mass map does not yet fully capture the true distribution of mass of the cluster at large radii.

\section{Discussion}
\label{sec:discussion}
The results presented in this work show the extraordinary potential of \Euclid\ to understand ICL formation. Below, we discuss the implications of our findings. 

\subsection{Formation process of the ICL of A\,2390}
\label{sec:formation}
The ICL records most of the relevant events that happen during the cluster's formation. Studying the stellar properties of the ICL provides information on the timing of the formation and insights into the processes that drove the build-up of the ICL and of the cluster itself. Comparing the ICL properties with those of the cluster members provides clues about the progenitors of this light.

Section~\ref{subsec:morph} shows that the population of elliptical galaxies dominates the cluster's core, and follows the distribution of the ICL very well. This similarity suggests that these two components are linked and may have a common origin. The analysis of substructures reveals the presence of two main galaxy groups besides the central core of the cluster. Indeed, we detect ICL in the north-western ICL sub-halo, overlapping with a small group of red galaxies, and also within the south-eastern group. These are all signs that the ICL of 
A\,2390 has an important contribution from the diffuse light of these smaller groups, the so-called pre-processing. This will be discussed in more detail in Sect.~\ref{sec:preprocessing}.

Radial colour gradients also provide relevant information about the physical processes that have formed or are forming the ICL and, consequently, the BCG. In Fig.~\ref{fig:radial_color}, we presented the \YE - \HE radial colour profiles of BCG+ICL for the three methods used in this work, namely \texttt{CICLE}, \texttt{DAWIS}, and the 1-D profile method (Ahad). The colour profile shows a flat behaviour in the inner $5$\,kpc turning into a negative slope until around 450\,kpc, becoming bluer with radius.

The central flatness suggests that the stellar populations of the inner $5$\,kpc are well mixed. This could be the case of a core \citep[like Abell 85, ][]{Montes2021}, a major merging event \citep[e.g.,][]{MT22} or caused by the presence of H$\alpha$ emission \citep{Alcorn2023}. 
We discard the possibility that the cause of the flat colour is a core, since that would appear as a flat region of the same size in the surface brightness profiles in Fig.~\ref{fig:SB_a2390}. However, the surface brightness profiles do not show this behaviour. Galaxies undergoing a major merger event could also produce a flat colour in the inner parts of the BCG+ICL profile \citep[e.g.,][]{DeMaio2018, MT22}. The second brightest member of the cluster is located $120$\,kpc north-west from the BCG and there is no massive galaxy in the very central region ($r<10$\,kpc). There is also no evidence of plumes or other signs of interaction between these two galaxies in the \Euclid images \citep[see, for example,][]{Martin2022}. Alternatively, this flat colour could indicate a past major merger \citep[see the cluster AS1063 in][]{MT18}.

%\mireia{i really need to see the west/east profiles to discuss this in detail}
%(\paola{should we add a map in the appendix?}\mireia{no no, this is not our work})
H$\alpha$ emission has been detected in the BCG of A\,2390 \citep{Alcorn2023}. This emission  is produced by a cone, extending to the north-west direction of $15.9$\,kpc of length (but not aligned with the north-western subgroup). \citet{Alcorn2023} argue that star formation could be happening in this H$\alpha$ cone, although more data are needed to support that. The \IE image (the highest resolution) shows an elongation similar to that of the HST imaging \citep{Alcorn2023}. This emission is also seen in Spitzer ($3.6$ and $8\,{\rm \mu m}$ imaging), and probably causes a flat area at $<$\,5\,kpc in the surface brightness radial profiles. 

From 5 to approximately 50\,kpc, we see a negative gradient in colour. Previous works reveal that the BCG+ICL profiles show negative colour gradients \citep[][]{Zibetti2005, MT14, Mihos2017,Iodice2017,Montes2021, golden-marx2023, zhang2024}, indicating gradients in their stellar populations, generally metallicity \citep[e.g.,][]{Peletier1990, Coccato2010, LaBarbera2012, Huang2018, Santucci2020}.

Figure~\ref{fig:vazdekis} shows the predicted colours of the \Euclid\ NIR filters for the \citet{Vazdekis2016} simple stellar population models, as a function of age and colour coded by metallicity. The models show that at a given metallicity, the colour is constant for most ages ($>2$\,Gyr). Therefore, the observed colour gradient beyond $5$\,kpc is indicative of a gradient in metallicity rather than age; from slightly supersolar metallicity at $\lesssim10$\,kpc to approximately [Fe/H] $\approx -0.7$ at 200--300\,kpc, in the ICL region. This trend is consistent with other metallicity determinations for the ICL previously reported in the literature \citep{MT14,Edwards2016, Edwards2020, MT18, Gu2020}. Note that observations show that at intermediate redshifts there are age gradients as well in BCG+ICL systems \citep{Toledo2011,MT14, Morishita2017, MT18}. Unfortunately, we cannot test here whether there is a decrease in age, since our colours are mostly sensitive to metallicity.

The mass-metallicity relationship of \citet{Gallazzi2005} can be used as a proxy to derive the mass of the progenitors of the ICL. A galaxy of the metallicity found here ([Fe/H] $\approx -0.7$ at 200--300\,kpc) has a mass of $\logten(M/\si{\solarmass})\approx 9.5$. If we take into account that it is the outer regions of the galaxies that are more easily stripped and that galaxies exhibit metallicity gradients, the ICL of A\,2390 could be produced by more massive galaxies, around $\logten(M/\si{\solarmass})\approx 10$.

Therefore, it seems that the ICL in A\,2390 (out to around 400\,kpc) is being built by the accretion of small systems with $\logten(M/\si{\solarmass})\approx 9.5$, or from the outskirts of galaxies similar to the Milky-Way, with $\logten(M/\si{\solarmass})\approx 10$. This agrees with previous estimates in other clusters \citep{MT14, MT18, DeMaio2018, MT22} and with simulations \citep{Purcell2007, Cui2014, Contini2014, Contini2019, Brown2024}. 
%\jesse{Does this agree with Perseus?}\mireia{mmm not with the errors we have now.}

\subsection{Using the ICL to constrain the mass distribution of 
A\,2390}

%It has been proposed that since ICL stars are not bound to galaxies, they should distribute themselves in the cluster's potential well in a similar way to dark matter. Consequently, 
The ICL appears as an excellent tracer of the total mass distribution of clusters of galaxies \citep{MT19, Alonso-Asensio2020, Sampaio-Santos2021, Yoo2024}. Since dark matter is the dominant component in galaxy clusters \citep[e.g.,][]{Carlberg1997}, tracing the total distribution of mass (baryons and dark matter) means tracing the distribution of dark matter. 
This property of the ICL has already been used to improve gravitational lensing maps by pointing out hidden mass substructures \citep[][]{Mahler2023}. In this work, taking advantage of \Euclid's abilities and ancillary data, we have explored the similarities between the ICL, mass, and X-rays. In Sect.~\ref{sec:mass_xrays}, we compared the 2-dimensional distribution of the BCG+ICL from \texttt{DAWIS}, the total mass, and the X-rays. 

The BCG+ICL maps for \texttt{CICLE} and \texttt{DAWIS} in all \Euclid NIR bands show a similar shape and orientation: elongated along a north-west to south-east axis with an ellipticity of $\sim$0.51. From Fig.~\ref{fig:morph_contours}, we can see that this is also the preferential axis along which most member galaxies are located. %However, Fig.~\ref{fig:mass_xrays} shows that this is not the case for the distribution of mass. 

The mass distribution in the inner parts is similar to the one of the BCG+ICL. However, beyond 300\,kpc, the mass contours become rounder than the diffuse light (ellipticity of 0.35 for the mass map, compared to 0.58 for the BCG+ICL), although still aligned with the main axis of the cluster. This alignment means that while the mass map `sees' the presence of the mass overdensities of the south-east and north-west subgroups of galaxies, it does not yet fully reproduce the observed substructure present in the BCG+ICL and cluster members. 
While strong lensing is more precise in the inner regions of a cluster, we rely on weak lensing in the outer parts, which is inherently noisy and sensitive to structures along the line of sight \citep{Hoekstra2001}. 
Similarly, the X-ray emission map is similar to the BCG+ICL and mass distributions in the inner parts ($\lesssim200$ kpc). 
In the outskirts, the X-ray emission follows the same north-west to south-east axis, although it is rounder than the other two distributions (ellipticity of 0.08). However, this could be due to a lower sensitivity problem preventing us from seeing the X-ray emission from the subgroups. %Alternatively, the X-rays relaxation times might be faster than that of the galaxies, taking on a morphology closer to that of a relaxed state while the galaxies are still interacting. 

\citet{Mahler2023} used the ICL to identify concentrations of mass that were not identifiable using X-rays, overdensities of members, or other means. This resulted in more accurate mass maps (RMS from 1\farcs26 to 0\farcs32). Given the improvement seen when including the ICL, we propose to use the ICL of 
A\,2390 to improve the mass maps, combining the presence of this light with the weak lensing at these radii, where the uncertainty of the mass maps is higher. 

\subsection{The dynamical state of A\,2390}

The ICL -- its shape, its content, and its stellar populations -- is intimately linked to the dynamical state of the cluster \citep[e.g.,][]{Jimenez-Teja2018, ContrerasSantos2024}. Therefore, studying the ICL of a cluster in detail can give us additional clues about its dynamical state.
For example, \citet{Kluge2024} used the spatial offset of the ICL centroid, combined with asymmetries in the X-ray distribution, to argue that the Perseus cluster, otherwise thought to be a relaxed cluster, is experiencing a merger with a group or small cluster. 
In the same way, here we discuss the dynamical state of A\,2390 using the clues we have collected throughout the paper. 

A\,2390 is a strong cool-core cluster believed to be dynamically relaxed \citep{Sonkamble2015}. However, the distribution of ICL and cluster members of A\,2390 reveal significant substructures (Figs.~\ref{fig:densmap}, \ref{fig:morph_contours}), which points to a complex mass distribution. As discussed in Sect. \ref{subsec:morph}, the peak of the distribution of elliptical galaxies is offset from the BCG by about 70\,kpc westward. This offset is caused by the presence of the two subgroups of galaxies (south-east and north-west), pushing the overall centroid of elliptical galaxies off the BCG. 

In contrast, the X-ray emission exhibits a single peak which coincides with the BCG. It also shows a slight asymmetry towards the south-east at about 300\,kpc (orange contours in Fig.~\ref{fig:mass_xrays}). On larger scales, the X-ray emission becomes more circular. 
Similarly, the peak of the gravitational lensing mass distribution (blue contours in Fig.~\ref{fig:mass_xrays}) coincides with the X-ray and BCG+ICL. 
The mass contours also appear to be more elongated at large radii, aligning more with the south-east to north-west axis of the cluster members and the ICL, although rounder than them. 
Consequently, our combined analysis of X-ray, total mass, galaxy members, and ICL suggests that A\,2390 is not going through any merger at this time. 
However, there is an elongation of both cluster members and ICL, caused by the presence of several subgroups of galaxies. %two subgroups of galaxies, north-west and south-east of the main or BCG group. 
The fact that the X-rays appear more relaxed than the ICL and the galaxy distribution agrees with the expected relaxation times for each component. The relaxation time of the X-ray emitting gas is of the order of hundreds of Myrs, while it is of the order of a few Gyrs for the galaxies in the cluster \citep{Sarazin1986,Binney2008}.

A\,2390 has relatively cool, dense gas that exists beyond the outer edge of the present-day cooling flow. \citet{Allen2001} proposed that this could be explained by dense gas from the cores of infalling subclusters, stripped and deposited in the core of the main cluster without strong shocking \citep{fabian1991}. Both south-eastern and north-western subgroups might be the remnants of substructures that merged in the past and provided their own cold gas, feeding the cooling flow of A\,2390 \citep{Alcorn2023}. 

\citet{Dutta2024} recently derived a weak lensing map of 
A\,2390. They found that the cluster is in a late-merger state, in agreement with our findings. They found a mass excess in the south-east at 220\,kpc from the BCG, coincident with the south-east group identified here and claim that this group is currently merging with the central core of the cluster. 
Additionally, we note that while we do not find evidence that 
A\,2390 is undergoing a merger at this time, \citet{Abraham1996} found a group at 650\,\arcsec (2.3\,Mpc) to the north-west that might also be falling into the cluster.

\subsection{An example of pre-processing in clusters of galaxies?}\label{sec:preprocessing}

Putting together all the information presented in this paper, diffuse light, cluster member galaxies, X-rays, and total mass distribution, then form a coherent picture of A\,2390. 
This system appears to be a missing link in the history of cluster assembly. In particular, the latest stages of merging of subgroups into the main cluster core. While we see no signs of perturbations consistent with the system undergoing a merger (see section above), there are substructures in the galaxies (especially in the spatial distribution of spirals, see Fig. \ref{fig:morph_contours}) and the ICL that suggest that the system may be one step away from relaxation. 

In addition, A\,2390 presents an extraordinary opportunity to constrain the `pre-processing channel' in ICL formation \citep{Mihos2004}. This pre-processing of the ICL may be an important channel for increasing the amount of this light as clusters assemble \citep{Ragusa2023}. The interaction of galaxies in the groups produces intragroup light (IGL) that will eventually become part of the ICL of the main cluster when the subgroups finally merge into the core.

%This suggests that this secondary group has yet to infall into the cluster. 
%which is suggestive of an ancient ICL that has not suffered any recent interaction with an infalling group. Contrarily, the ICL associated with group 3 is blue, which indicates that it must be intragroup light that have not mixed yet with the diffuse light of the cluster. In an intermediate stage, the ICL attached to groups number 5 and 6 is green, which implies some degree of mixing with the extant ICL population. This also suggests that the southeastern structure is dynamically more evolved than the northwestern one and their groups associated either have already crossed the core of the cluster or are in an ongoing merger infalling from the southeast. Their mean velocities show that both groups are in the background of the cluster, although the high velocity dispersion indicate that they could be consistent with that of the cluster. \paola{why don't you move this discussion to the discussion section? maybe in 6.4?}
%\mireia{I'll take a look and move it.}

The position of the two subgroups, south-eastern and north-western, mimics the filamentary structure around A\,2390 \citep[beyond its virial radius, see][]{Abraham1996, Li2009}. The mean velocities and velocity dispersions of both groups indicate that they are consistent with that of the main cluster core. Therefore, what we are witnessing is groups of galaxies that have infallen into the A\,2390 halo in the past, through cosmic web filaments, and are finally reaching the inner parts of the cluster. Eventually, these groups will merge with the main core, bringing with them pre-processed IGL and new galaxies that will build up the BCG+ICL system. The IGL contribution from these groups could also account for the high fraction of ICL over BCG+ICL that we find in Sect.~\ref{sec:fractions}. 

An additional clue is that the IGL brought by these subgroups is not well-mixed, a telltale sign of the early stages of the merging process. The IGL around the north-western group shows both red and blue colours (Fig.~\ref{fig:rgb_groups}), indicating that the IGL of this group itself has not entirely been mixed, so this group may be in a pre-merger stage. 

The IGL attached to the south-eastern structure shows a more intermediate colour, which implies some degree of mixing with the extant ICL population. This mixing suggests that the south-eastern structure might be dynamically more evolved than the north-western one, and their associated groups either have already crossed the core of the cluster or are in an ongoing merger infalling from the south-east. This scenario could be consistent with the orbits for a few emission-line galaxies that belong to both substructures of A\,2390 found by \cite{Liu2021}.

The north-western group contributes 21\,\% to the total amount of ICL in the inner $300$\,kpc of the cluster, while the south-eastern group contributes 9\,$\%$, giving a total flux fraction of 30\,$\%$. \citet{Contini2024} showed that around 20\,$\%$ of the ICL in simulations comes from pre-processing, which agrees with the estimates presented here. When these subgroups merge into the main core, they will contribute to around 30\,$\%$ of the ICL. 
Note that this estimate does not take into account the ICL that will be produced when the galaxies interact, and should therefore be regarded as a lower limit.

\section{Conclusions}
\label{sec:conclusions}

We have presented an in-depth analysis of the intracluster light of the A\,2390 cluster of galaxies using the ERO imaging taken by \Euclid.
To enable the analysis of the ICL we removed contamination by the Galactic cirrus by applying a novel method that relies on the different scales of the Galactic dust and the ICL. We specifically used the wavelet-based \texttt{DAWIS} code to derive a dust map to remove the contribution of the cirrus from the NIR \Euclid images.  

We have detected the ICL out to 600\,kpc. We found that the colour of the BCG+ICL is constant for the first $5$\,kpc, probably caused by the presence of an H$\alpha$ emission cone, followed by a decline (negative gradient) until about 450\,kpc. This gradient is associated with a gradient in metallicity from slightly supersolar metallicity to [Fe/H]\,$\approx -0.7$ in the ICL region ($>100$\,kpc). This value suggests that the ICL is being built by satellites of mass $\logten(M/\si{\solarmass})\approx10$. 

We compared the distribution of member galaxies with that of the ICL. The ICL agrees very well with the distribution of member galaxies, especially the cluster's elliptical galaxies. We find that there are ICL patches associated with subgroups of galaxies in the cluster, suggesting that the pre-processing of the ICL in these groups (a lower limit of $\sim 30\,\%$ of the total ICL) is an important mechanism of ICL production.

We also compare the 2D distributions of diffuse light, total mass (from gravitational lensing), and X-rays. While the three distributions are quite similar in the inner regions ($<300$\,kpc), they start to diverge further out. It seems that the mass map is not yet able to fully capture the substructure seen in A\,2390 in the ICL and the member galaxies. 

The results presented here demonstrate the potential of \Euclid for understanding the formation of the ICL and the assembly history of galaxy clusters. It also highlights the importance of having large samples to fully explore the different ICL formation mechanisms and their contributions.

%%%%%%%%%%%%%%%
\begin{acknowledgements}

\AckEC  

\AckERO

This research made use of Photutils, an Astropy package for detection and photometry of astronomical sources \citep{Bradley2024}. This work was partly done using GNU Astronomy Utilities (Gnuastro, ascl.net/1801.009) version 0.21; work on Gnuastro has been funded by the Japanese Ministry of Education, Culture, Sports, Science, and Technology (MEXT) scholarship and its Grant-in-Aid for Scientific Research (21244012, 24253003), the European Research Council (ERC) advanced grant 339659-MUSICOS, the Spanish Ministry of Economy and Competitiveness (MINECO, grant number AYA2016-76219-P) and the NextGenerationEU grant through the Recovery and Resilience Facility project ICTS-MRR-2021-03-CEFCA.
M.M. acknowledges support from the project PCI2021-122072-2B, financed by MICIN/AEI/10.13039/501100011033, and the European Union “NextGenerationEU”/RTRP, from the grant RYC2022-036949-I financed by the MICIU/AEI/10.13039/501100011033 and by ESF+ and program Unidad de Excelencia Mar\'{i}a de Maeztu CEX2020-001058-M.

A. E. acknowledges fundings by the CNES post-doctoral fellowship program.
F.D. acknowledges long-term support from CNES. 
P.D. acknowledges funding from the Italian INAF Large Grant 12-2022.
\end{acknowledgements}

%%%%%%%%%%%%%%%
% Here comes the reference list, generated via bibtex from
% your bibfile my.bib and Euclid.bib. Please make sure that
% the same paper is not referenced twice, one from your my.bib
% file, and once from Euclid.bib.
%

\bibliography{merged_bib}
\appendix

\section{WISE dust maps} \label{app:cirri}

\begin{figure*}[h!]
\begin{center}
\includegraphics[width=\textwidth]{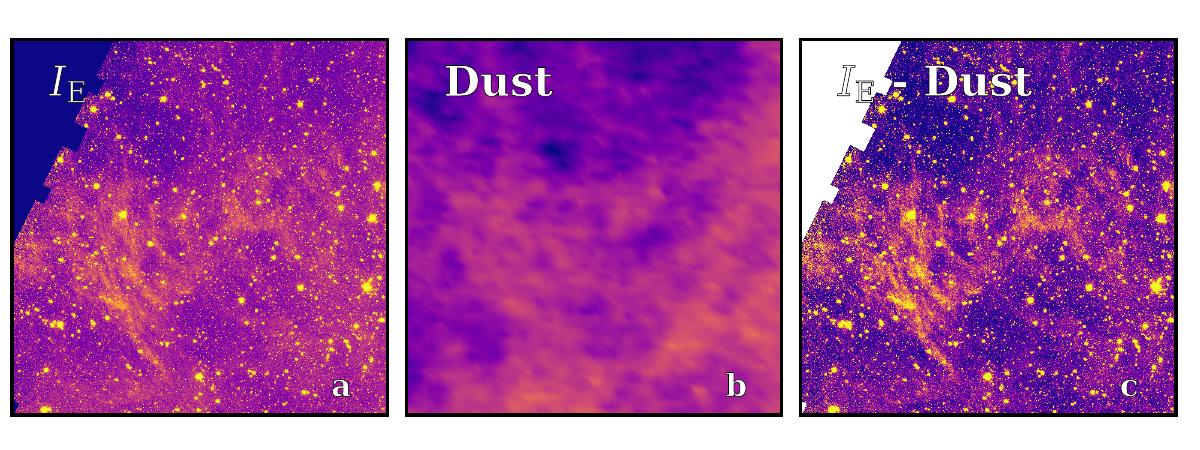}
\end{center}
\caption{ Example of the problem of using the WISE dust maps to correct for cirrus in the  \Euclid\ images. Panel a shows a region of $25\arcmin\times25\arcmin$ in the \IE\ band. Panel b shows the corresponding WISE dust map for the region. Panel c shows the result of subtracting the WISE dust map from the \IE\ image. The residual shows cirrus filaments that the WISE map does not capture.}
\label{fig:wise_vis}
\end{figure*}

Far-infrared maps are commonly used to compare or even remove cirrus in the optical bands \citep[e.g.,][]{Mihos2017, Kluge2020, Kluge2024}. As mentioned in Sect.~\ref{sec:wise}, we approached the cirrus removal in the \Euclid FoV of A\,2390 using the WISE 12\,\micron\ dust emission maps from \citet[][]{Meisner2014}. Figure \ref{fig:wise_vis} shows a region of the \Euclid's A\,2390 FoV centred on RA = \ra{21;54;44.0}, Dec = \ang{+17;41;4.6}, to illustrate the methodology. The main problem of this approach is the use of dust maps with a significantly lower spatial resolution (26\,\arcsec) than in \Euclid ($<1$\,\arcsec) and, therefore, it cannot provide the level of detail required to properly subtract the cirrus. However, the fact that there are unmatched spatial features between the FIR and the optical/NIR could also play a role here. As an example, the large-scale pockets of bright dust at the bottom of the WISE map in Fig.~\ref{fig:wise_vis} are not seen in the \Euclid\ bands (see also Sect.~\ref{sec:wise} and Fig.~\ref{fig:cirrus_rgb_VIS_Y_J}).

\section{CICLE and \texttt{DAWIS} BCG+ICL maps}

\begin{figure*}
\begin{center}
\includegraphics[width=0.85\textwidth]{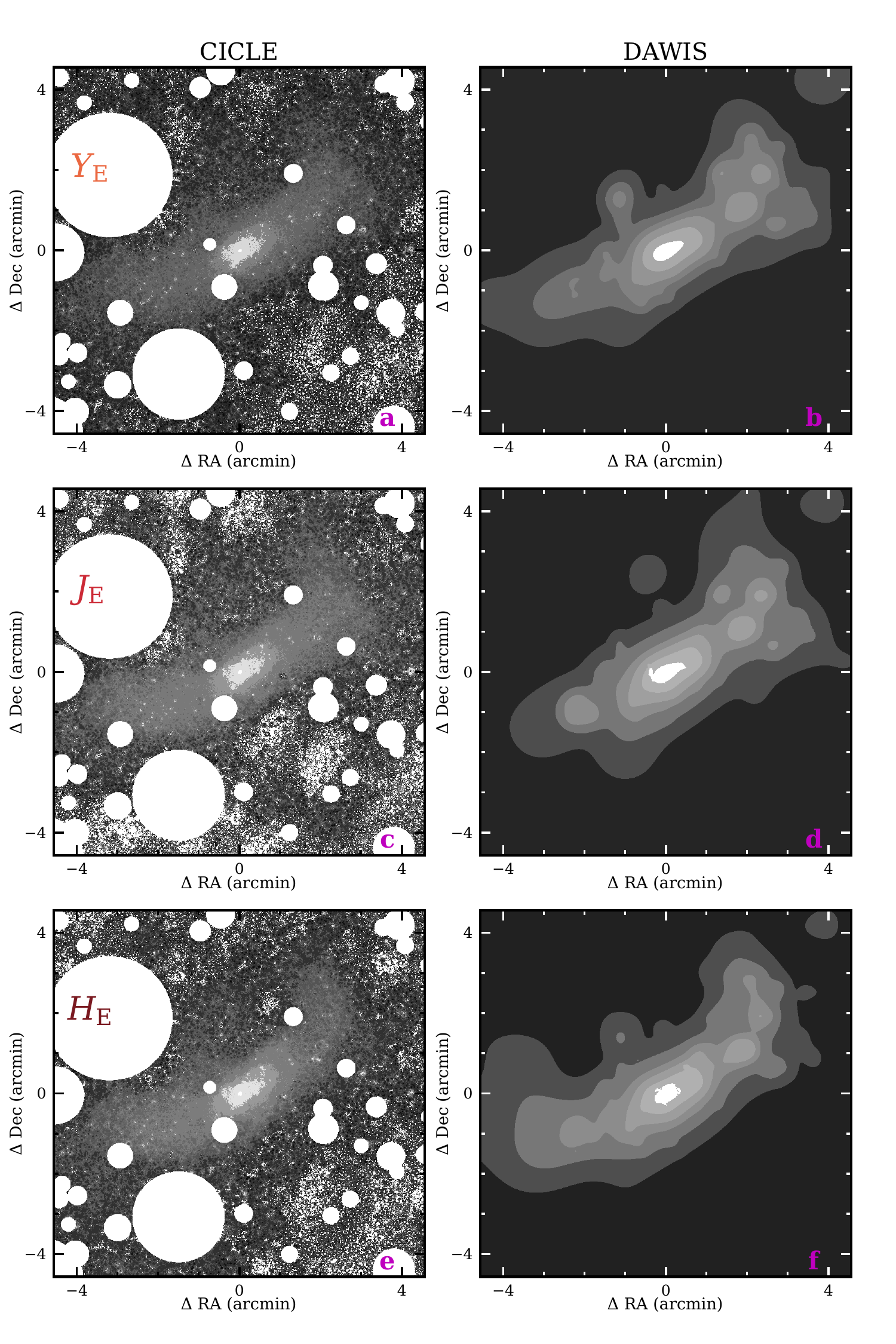}
\end{center}
\caption{BCG+ICL maps for the NIR \Euclid's \YE, \JE, and \HE bands for \texttt{CICLE} (panels a, c, and e), and \texttt{DAWIS} (panels b, d, and f). }
\label{fig:icl_maps}
\end{figure*}

Figure~\ref{fig:icl_maps} presents the \texttt{CICLE} (left panels) and \texttt{DAWIS} (right panels) BCG+ICL maps, for each NISP filter, \YE\ (upper panels), \JE\ (middle panels), and \HE (lower panels).

%\section{MW Extinction maps}
%\begin{figure}
%\begin{center}
%\includegraphics[width=0.5\textwidth]{figures/Dust_map_Y-1.png} 
%\includegraphics[width=0.5\textwidth]{figures/Dust_map_J-1.png}
%\includegraphics[width=0.5\textwidth]{figures/Dust_map_H-1.png}
%\end{center}
%\caption{Milky Way extinction maps. Just for discussion purposes.}
%\label{fig:MW_extinction}
%\end{figure}

\section{Colours of cluster members}\label{app:members}

\begin{figure}
\begin{center}
\includegraphics[width=0.5\textwidth]{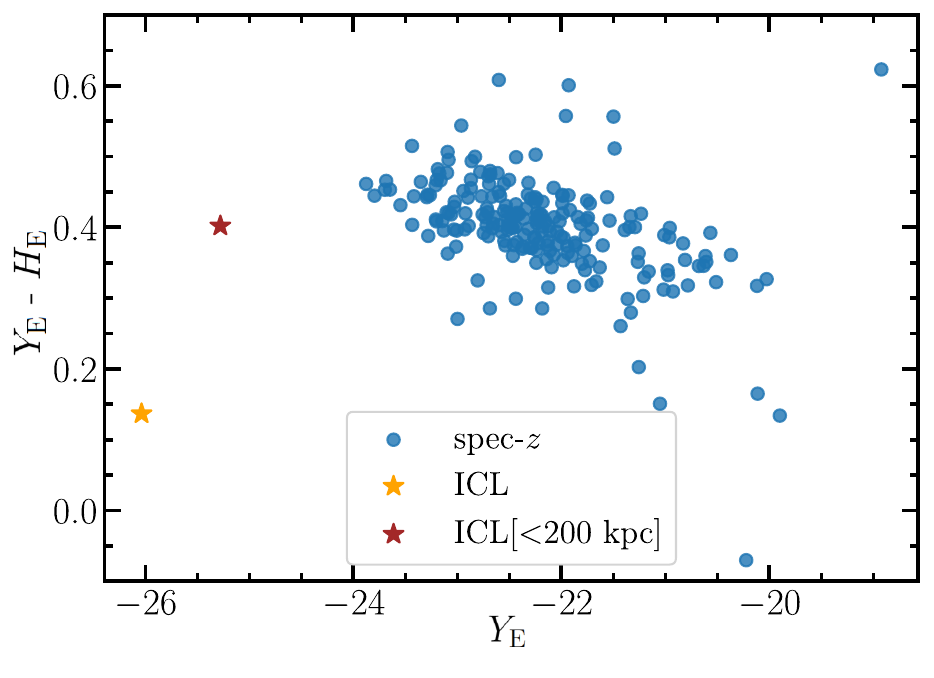}
\end{center}
\caption{Colour-magnitude diagram for A\,2390 galaxies. Red and gold stars represent the total and the core ($r<200$\,kpc) ICL colour.}
\label{fig:color_histogram}
\end{figure}

Figure \ref{fig:color_histogram} shows the $\YE-\JE$ and $\JE-\HE$ colour distribution of cluster member galaxies. The colours are taken from the catalogue shown in Sect.~\ref{sec:members}.

\begin{figure}
\begin{center}
\includegraphics[width=0.4\textwidth]{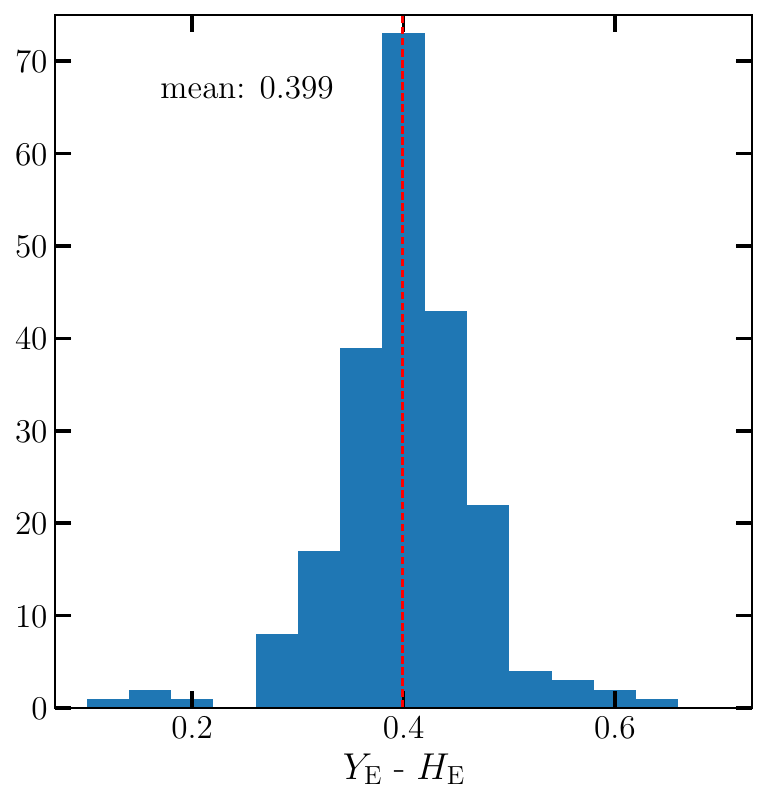}
\end{center}
\caption{Colour distribution of cluster members.}
\label{fig:CMD}
\end{figure}

\section{Stellar population models in the \Euclid\ NIR filters}\label{app:models}

\begin{figure}
\begin{center}
\includegraphics[width=0.4\textwidth]{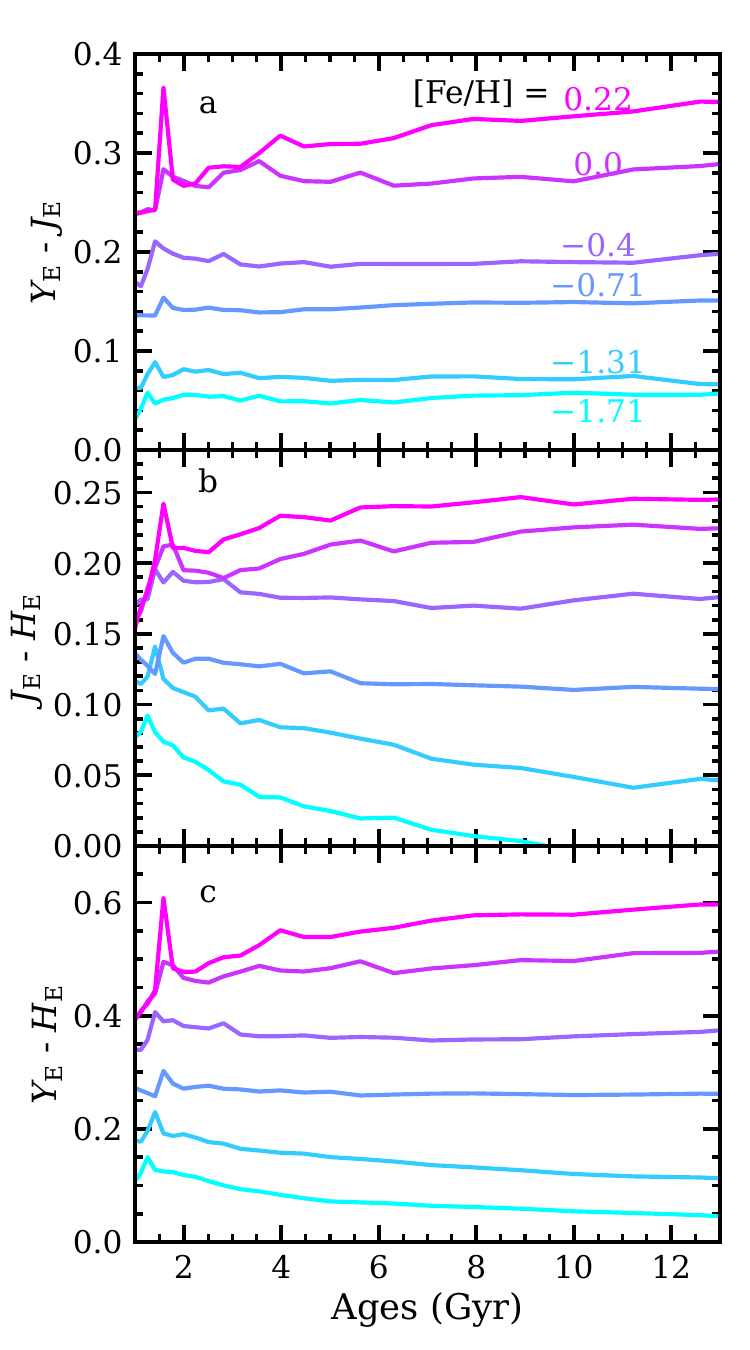}
\caption{\Euclid\  NIR colours of the \citet{Vazdekis2016} single stellar population models shown as a function of stellar age, for A\,2390. When the stellar population is older than 2\,Gyr, its $\YE-\JE$ (panel a), $\JE-\HE$ (panel b), and $\YE-\HE$ (panel c) colours are fairly constant with age but vary with metallicity.
\label{fig:vazdekis}} 
\end{center}

\end{figure}

Here, we explore how the \Euclid\ NIR colours of stellar populations vary with stellar age. In Fig.\,\ref{fig:vazdekis} we show the NIR colours of the single stellar population models from \citet{Vazdekis2016}, assuming a \citet{Kroupa2001} initial mass function, convolved with the \Euclid\ filters, following the prescriptions in \citet{Montes2014}. The models are redshifted to the redshift of A\,2390 ($z=\,0.228$). 
The models are colour-coded according to the metallicity, as labelled in the middle panel, and are shown as a function of stellar age.
The colour $\YE-\HE$, the one used in this paper, of the stellar populations shows almost no variation with age once the stellar ages are $>2$\,Gyr. Therefore, the $\YE-\HE$ colours in A\,2390 are mostly insensitive to age, so we are tracing changes in the metallicity of the stellar populations of the BCG+ICL.

\end{document}